\documentclass[twocolumn,showpacs,preprintnumbers,amsmath,amssymb]{revtex4}

\usepackage{graphicx}
\usepackage{dcolumn}
\usepackage{bm}

\begin{document}
\newcommand{\gdhi}{\ooalign{\hfil/\hfil\crcr$\partial$}}

\def\Sp{\mathop{\mathrm{Sp}}\nolimits}
\def\sgn{\mathop{\mathrm{sgn}}\nolimits}
\def\erfc{\mathop{\mathrm{erfc}}\nolimits}
\def\tr{\mathop{\mathrm{tr}}\nolimits}
\def\in{\mathop{\mathrm{ln}}\nolimits}
\def\as{\mathop{\mathrm{as}}\nolimits}
\def\val{\mathop{\mathrm{val}}\nolimits}


\title{Axially symmetric multi-baryon solutions and their quantization\\
in the chiral quark soliton model }

\author{S.Komori}
\author{N.Sawado}
\email{sawado@ph.noda.tus.ac.jp}
\author{N.Shiiki}
\email{norikoshiiki@mail.goo.ne.jp}
\affiliation{
Department of Physics, Faculty of Science and Technology, 
Tokyo University of Science, Noda, Chiba 278-8510, Japan 
}
\date{\today}

\begin{abstract}
In this paper, we study axially symmetric solutions with $B=2-5$ in 
the chiral quark soliton model. 
In the background of axially symmetric chiral fields, 
the quark eigenstates and profile functions of the 
chiral fields are computed self-consistently. The resultant
quark bound spectrum are doubly degenerate due to the symmetry 
of the chiral field. Upon quantization, various observable spectra 
of the chiral solitons are obtained. Taking account of 
the Finkelstein-Rubinstein constraints, we show that 
our results exactly coincide with the physical observations for 
$B=2$ and $4$ while $B=3$ and $5$ do not.
\end{abstract}


\maketitle

\section{\label{sec:level1}Introduction\protect\\ }
The chiral quark soliton model(CQSM) was developed in
1980's as a low-energy effective theory of QCD. 
Since it includes the Dirac sea quark 
contribution and explicit valence quark degrees of 
freedom, the model interpolates between the 
constituent quark model and the Skyrme 
model \cite{diakonov88,reinhardt88,meissner89,report,wakamatsu91}. 
The CQSM is derived from the instanton liquid model of
QCD vacuum and incorporates the non-perturbative
feature of the low-energy QCD, spontaneous chiral
symmetry breaking. 
It has been shown that 
the $B=1$ solution provides correct observables as a nucleon 
including mass, electromagnetic value, spin carried
by quarks, parton distributions and octet
SU(3) baryon spectra. 

For $B=2$, the stable axially symmetric
soliton solution was found in~\cite{sawado98}. 
The solution exhibits doubly degenerate bound spectrum 
of the quark orbits in the background of the axially symmetric  
chiral field with winding number two. Upon quantization, 
various dibaryon spectra were obtained, showing that the quantum 
number of the ground state exactly coincide with that of physical 
deuteron~\cite{sawado00,sawado02}. For $B> 2$, the Skyrme model 
predicts that the solutions have only discrete, 
crystal-like symmetries~\cite{braaten90,sutcliffe97,manton98}. 
According to the prediction, we studied the CQSM  
with $B=3$ tetrahedrally symmetric chiral fields 
and obtained triply degenerate spectrum of the quark orbits
~\cite{sawado02t}. Its large degeneracy indicates that 
the terahedrally symmetric solution may be the lowest-lying 
configuration. For $B>3$, one can also expect that the 
lowest-lying solutions in the CQSM 
inherits the discrete symmetries predicted in the Skyrme model. 
Studying solutions with those symmetries in CQSM 
is, however, formidable especially for quantization. 
Thus, before embarking those discrete symmetries, 
it will be instructive to study axially symmetric 
solutions which are much simpler. Besides, considering 
the fact that for some higher baryon numbers, the ground 
states of the skyrmions do not agree with the experimental 
observation~\cite{irwin} , the possibility that axially symmetric 
solutions provide correct ground states can not be 
excluded. Recently, it was found in Ref.~\cite{grigoriev} that
in the BPS monopoles all axially symmetric solutions up to 
chrage five have lower energy than that of discrete symmetries. 
A cylindrical shape isomer in $^{12}C$  
was also found in the Skyrme model framework~\cite{nikolaev}.
Research of axially symmetric solitons is thus in progress.

In Sec.\ref{sec:level2}, we shall obtain axially symmetric 
classical soliton solutions with $B=2-5$. The solutions exhibit doubly 
degenerate spectra due to their axial symmetry.
Such degeneracy generates large shell gaps and confirms  
that the solutions are stable local minima. 
In Sec.\ref{sec:level3}, we shall quantize the obtained classical 
solitons semiclassically. Imposing the Finkelstein-Rubinstein 
constraints on the states, the ground states of the axially symmetric 
solitons are constructed and examined if they agree with the 
experimental observation. 
In Sec.\ref{sec:level4} is the detail of the numerical analysis 
used to obtain the classical and quantum solutions.
Conclusions and discussions are in Sec.\ref{sec:level5}. 
 
\section{\label{sec:level2}Axially Symmetric Classical Solutions}
The CQSM is derived from the instanton liquid model of 
the QCD vacuum and incorporates the nonperturbative feature
of the low-energy QCD, spontaneous chiral symmetry breaking. 
The vacuum functional is defined by \cite{diakonov88}
\begin{eqnarray}
	{\cal Z} = \int {\cal D}\pi{\cal D}\psi{\cal D}\psi^{\dagger}\exp \left[ 
	i \int d^{4}x \, \bar{\psi} \left(i\!\!\not\!\partial
	- MU^{\gamma_{5}}\right) \psi \right]	 \label{vacuum_functional}
\end{eqnarray} 
where the SU(2) matrix
\begin{eqnarray}
	U^{\gamma_{5}}= \frac{1+\gamma_{5}}{2} U + \frac{1-\gamma_{5}}{2} U^{\dagger} 
	\,\,\,{\rm with} \,\,\,\,
	U=\exp \left( i \bm{\tau} \!\cdot\! \bm{\pi}/f_{\pi} \right) \nonumber
\end{eqnarray}
describes chiral fields, $\psi$ is quark fields and $M$ is the constituent 
quark mass. $f_{\pi} $ is the pion decay constant and experimentally 
$f_{\pi} \sim 93 {\rm MeV}$. 

The $B=1$ soliton solution has been studied in detail at classical and 
quantum level in \cite{diakonov88,reinhardt88,meissner89,
report,wakamatsu91}.  
To obtain solutions with $B>1$, we shall employ the chiral fields with 
winding number $B$ in the Skyrme Model as the background of quarks, 
which can be justified as follows. 

In Eq.(\ref{vacuum_functional}), performing the functional integral over 
$\psi$ and $\psi^\dagger$ fields, one obtains the effective action
\begin{equation}
	S_{\rm eff}(U)=-iN_{c} \Sp \ln iD = -iN_c \log \det iD, 
	\label{effective_action2}
\end{equation}
where $iD=i \gdhi - M U^{\gamma_5}$ is the Dirac operator.
The classical solutions can be obtained by the extremum condition 
of (\ref{effective_action2}) with respect to $U$. 
For this purpose, let us consider the derivative expansion of 
the action \cite{dhar85,ebert86,wakamatsu91}. 
Up to quartic terms, we have,  
\begin{eqnarray}
	&&S_{\rm eff}=\int d^4x \biggl[ C {\rm tr}(L_\mu L^\mu)
	 \nonumber \\
	&&+\frac{N_c}{32\pi^2}
	{\rm tr}\Bigl\{\frac{1}{12}[L_\mu,L_\nu]^2-\frac{1}{3}(\partial_\mu L^\mu)
	+\frac{1}{6}(L_\mu L^\mu)^2 \Bigr\} \biggr], \nonumber \\
	\label{expand_action}
\end{eqnarray}
where $L_\mu=\partial_\mu U U^\dagger$. Suitably adjusting the coefficients, 
one can identify (\ref{expand_action}) with the Skyrme model action. 
Therefore, it will be justified to adopt the configurations of the solutions 
in the Skyrme model to chiral fields in the CQSM. 

In the Skyrme model the minimal energy pion configuration 
with $B=2$ has an axial symmetry \cite{manton87} and can be 
written by 
\begin{equation}
	U(\bm{x})= \cos F(\rho,z)+i\bm{\tau} \cdot \hat{\bm{n}} \sin F(\rho,z),
	\label{ansatz}
\end{equation}
where
\begin{equation}
	\hat{\bm{n}}=(\sin \Theta(\rho,z) \cos m_{{\rm w}} \varphi,\sin \Theta(\rho,z) 
	\sin m_{{\rm w}} \varphi,\cos \Theta(\rho,z))
	\label{ansatz2}
\end{equation}
and $m_{{\rm w}}$ is the winding number of the pion fields. We shall use this 
configuration in the backgound to obtain axially symmetric chiral quark 
solitons.

In the CQSM, the number of valence quark is associated with 
the baryon number such that the baryon number $B$ soliton consist of 
$N_c\times B$ valence quarks. 
If the correlation between quarks is sufficiently strong, 
their binding energy become large and the valence quarks 
can not be observed as positive energy particles
~\cite{kahana84,balachandran98}.
Thus, one gets the picture of the topological soliton model 
in the sense that the baryon number coincide with the winding 
number of the background chiral field when the valence quarks 
occupy all the levels diving into negative energy region. 

Let us rewrite the effective action in (\ref{effective_action2}) as 
\begin{eqnarray}
	S_{\rm eff}= -iN_c \log \det(i \gdhi - M U^{\gamma_5}) \nonumber \\
	=-iN_c\log \det\bigl(i\partial_t-H(U^{\gamma_5})\bigr) \label{effective_det}
\end{eqnarray}
where 
\begin{eqnarray}
	H(U^{\gamma_5})=-i\alpha\cdot\nabla + \beta MU^{\gamma_{5}}\,.
	\label{hamiltonian}
\end{eqnarray}
The classical energy of the soliton can be estimated from the quark determinant
in Eq.(\ref{effective_det}) \cite{rajaraman,reinhardt89}. 
We introduce the eigenstates of operatos, 
$i\partial_t-H(U^{\gamma_5})$ and $H(U^{\gamma_5})$, such that 
\begin{eqnarray}
	&&H(U^{\gamma_5})\phi_{\mu}(\bm{x})=E_{\mu}\phi_{\mu}(\bm{x}) \label{eigen_h}\\
	&&\bigl(i\partial_t-H(U^{\gamma_5})\bigr)\Psi_{\mu,n}=\lambda_{\mu,n}\Psi_{\mu,n}\, .
	\label{eigenequation}
\end{eqnarray}
where $\Psi_{\mu,n}=e^{-i\omega_{n}t}\phi_{\mu}$ and 
$\lambda_{\mu,n}=-E_\mu+\omega_n$.
Imposing on $\Psi_{\mu,n}$ the anti-periodicity condition,  
$\Psi_{\mu,n}(\bm{x},T)=-\Psi_{\mu,n}(\bm{x},0)$, 
reads  
\begin{eqnarray}
\omega_nT=(2n+1)\pi. 
\end{eqnarray}
The determinant in Eq.(\ref{effective_det}) then becomes 
\begin{eqnarray}
	\det(i\partial_t-H)&=&\prod_{\mu,n}\lambda_{\mu,n} \nonumber \\
	&=&\prod_{\mu,n}\Bigl(-E_\mu+\frac{(2n+1)\pi}{T}\Bigr) \nonumber \\
	&=&C\prod_{\mu,n\ge0}\Bigl(1-\frac{|E_\mu|^2 T^2}{(2n+1)^2\pi^2} \Bigr) \nonumber \\
	&=&C\prod_{\mu}\cos\Bigl(\frac{1}{2}|E_\mu|T\Bigr) \nonumber \\
	&=&\frac{C}{2}\exp\Bigl(i\frac{1}{2}\sum_\mu|E_\mu|T\Bigr) \nonumber \\
	&&\times\prod_\mu\Bigl(1+\exp(-i|E_\mu|T)\Bigr) \label{det}
\end{eqnarray}
where 
\begin{equation}
	C=\prod_{n\ge0}\Bigl(-\frac{(2n+1)^2\pi^2}{T^2}\Bigr) \nonumber
\end{equation}
and the product formula for the cosine function 
$\cos(z)=\prod^\infty_{n\ge1}(1-4z^2/(2n-1)^2\pi^2)$ has been used.
Inserting (\ref{det}) into (\ref{effective_det}), one obtains 
\begin{eqnarray}
	S_{\rm eff}=-N_cT\sum_\mu n_\mu|E_\mu|+N_cT\frac{1}{2}\sum_\mu |E_\mu| 
\end{eqnarray}
where $n_\mu$ is the valence quark occupation number which takes  
values only 0 or 1. 
Correspondingly, the classical energy is given by
\begin{eqnarray}
	E=E_{\rm val}+E_{\rm field}
\end{eqnarray}
where
\begin{eqnarray}
	&&E_{\rm val}=N_c\sum_\mu n_\mu |E_\mu|\,, \nonumber\\
	&&E_{\rm field}=-\frac{1}{2}N_c\sum_\mu |E_\mu|\,.\nonumber
\end{eqnarray}
representing the valence quark and 
sea quark contribution to the total energy respectively. 

The effective action $S_{\rm eff}(U)$ is ultraviolet divergent 
and hence must be regularized. 
Using the proper-time regularization scheme 
\cite{schwinger51}, we can write 
\begin{eqnarray}
	&&S^{{\rm reg}}_{{\rm eff}}[U]
	=\frac{i}{2}N_{c}
	\int^{\infty}_{1/\Lambda^2}\frac{d\tau}{\tau}{\rm Sp}\left(
	{\rm e}^{-D^{\dagger}D\tau}-{\rm e}^{-D_{0}^{\dagger}D_{0}\tau}\right) 
	\nonumber \\
	&&\hspace{12mm}=\frac{i}{2}N_{c}T\int^{\infty}_{-\infty}\frac{d\omega}{2\pi}
	\int^{\infty}_{1/\Lambda^2}\frac{d\tau}{\tau} \nonumber \\
	&&\hspace{20mm}\times{\rm Sp}\Bigl[{\rm e}^{-\tau (H^2+\omega^2)} 
	-{\rm e}^{-\tau (H_{0}^2+\omega^2)}\Bigr]
	\label{regularized_action}
\end{eqnarray} 
where $D_{0}$ and $H_{0}$ are operators with $U=1$.
The total energy is then given by
\begin{equation}
	E_{\rm static}[U]=E_{\rm val}[U]+E_{\rm field}[U]-E_{\rm field}[U=1]
\end{equation}
where
\begin{eqnarray}
	&&E_{\rm val}=N_c\sum_{i}E^{(i)}_{\rm val} \nonumber \\
	&&E_{\rm field}=N_c\sum_{\mu}\left\{ {\cal N}(E_\mu)|E_{\mu}|+\frac{\Lambda}
	{\sqrt{4\pi}}\exp \left[ - \left( \frac{E_{\mu}}{\Lambda} \right)^2 
	\right] \right\}\nonumber 
\end{eqnarray}
with
\begin{equation}
	{\cal N}(E_{\mu})= -\frac{1}{\sqrt{4\pi}}\Gamma \left(\frac{1}{2}, 
	\left(\frac{E_{\mu}}{\Lambda} \right)^2 \right) \,\nonumber
\end{equation}
and $E^{(i)}_{\rm val}$ is the valence energy of the $i$ th 
valence quark. $\Lambda$ is a cutoff parameter evaluated by
the condition that the derivative expansion of (\ref{regularized_action}) 
reproduces the pion kinetic term with the correct coefficient, {\it i.e.},
\begin{eqnarray}
	f_{\pi}^2=\frac{N_{c}M^2}{4\pi^2}\int^{\infty}_{1/\Lambda^2} 
	\frac{d\tau}{\tau}{\rm e}^{-\tau M^2}
	\, . \label{cutoff_parameter}
\end{eqnarray}
The extremum conditions for the total energy 
\begin{equation}
\frac{\delta}{\delta F(\rho,z) } E_{\rm static}[U]=0\,,~~~\frac{\delta}{\delta 
\Theta(\rho,z) } E_{\rm static}[U]=0
\end{equation}
yield the following equations of motion for the profile functions,
\begin{eqnarray}
	&&R^{T}(\rho,z)\cos \Theta (\rho,z)=R^{L}(\rho,z)\sin \Theta (\rho,z)
	\label{eq_proff}  \\
	&&S(\rho,z)\sin F (\rho,z)=P(\rho,z)\cos F (\rho,z)~\label{eq_proft}
	\end{eqnarray}
where
\begin{equation}
	P(\rho,z)=R^{T}(\rho,z)\sin \Theta (\rho,z)+R^{L}(\rho,z)\cos \Theta (\rho,z)\,.
	\label{eq_p}
\end{equation}
In terms of eigenfunction $\phi$ in Eq.(\ref{eigen_h}), 
$R^{T}$,$R^{L}$ and $S$ are given by 
\begin{eqnarray}
	&&R^{T}(\rho,z)=R^{T}_{\rm val}(\rho,z)+R^{T}_0(\rho,z) \\
	&&R^L(\rho,z)=R^{L}_{\rm val}(\rho,z)+R^L_{0}(\rho,z)   \\
	&&S(\rho,z)=S_{\rm val}(\rho,z)+S_{0}(\rho,z)
\end{eqnarray}
where
\begin{eqnarray}
	&&R^{T}_{\rm val }(\rho,z)=\sum_{i}\int d \varphi \bar{\phi}_i(\rho,\varphi,z)
	 i \gamma_5~\nonumber\\
	&&\hspace{2cm}\times
	(\tau_1 \cos m_{{\rm w}}\varphi + \tau_2 \sin m_{{\rm w}}\varphi ) 
	\phi_i(\rho,\varphi,z)\,, \nonumber\\
	&&R^{T}_{0}(\rho,z)= \sum_{\mu} {\cal N}(E_{\mu}) \sgn(E_{\mu}) \int d 
	\varphi \bar{\phi}_{\mu}(\rho,\varphi,z)
	 i \gamma_5\nonumber\\
	&&\hspace{2cm}\times
	(\tau_1 \cos m_{{\rm w}}\varphi + \tau_2 \sin m_{{\rm w}}\varphi ) 
	\phi_{\mu}(\rho,\varphi,z)\,, \nonumber\\
	&&R^L_{\rm val}(\rho,z)=\sum_{i}\int d \varphi \bar{\phi}_i(\rho,\varphi,z)i 
	\gamma_5\tau_3\phi_i(\rho,\varphi,z)\,, \nonumber\\
	&&R^L_{0}(\rho,z)= \sum_{\mu} {\cal N}(E_{\mu}) \sgn(E_{\mu}) \int d \varphi 
	\bar{\phi}_{\mu}(\rho,\varphi,z)
	 i \gamma_5 \nonumber\\
	&&\hspace{2cm}\times
	\tau_3 \phi_{\mu}(\rho,\varphi,z)\,, \nonumber\\
	&&S_{\rm val}(\rho,z)=\sum_{i}\int d \varphi \bar{\phi}_i(\rho,\varphi,z)
	\phi_i(\rho,\varphi,z)\,,  \nonumber\\
	&&S_{0}(\rho,z)= \sum_{\mu} {\cal N}(E_{\mu}) \sgn(E_{\mu}) \int d \varphi 
	\bar{\phi}_{\mu}(\rho,\varphi,z)\phi_{\mu}(\rho,\varphi,z)\,.\nonumber 
\end{eqnarray}
and subscripts, $0$ and val, represent the vacuum and valence quark 
contributions respectively.
The boundary conditions for the profile functions were constructed 
by Braaten and Carson~\cite{braaten88};
\begin{eqnarray}
	&&F(\rho ,z)\rightarrow 0 ~~\as~~ \rho^2+z^2\rightarrow \infty , \nonumber \\
	&&F(0,0)=-\pi, ~~~\Theta(0,z)=  \left\{ 
	\begin{array}{c} 0, ~~~ z>0  \\
	\pi, ~~~z<0  
	\end{array}\right.\;.
	\label{boundary}
\end{eqnarray}
The  procedure to obtain the self-consistent solutions of Eq.(\ref{eq_proff})
and (\ref{eq_proft}) is $1)$~solve the eigenequation in Eq.(\ref{eigenequation})
under assumed initial profile functions $F_0(\rho,z),\Theta_0(\rho,z)$ which 
satisfy the boundary conditions eqs.(\ref{boundary}), $2)$~use the resultant 
eigenfunctions and eigenvalues to calculate $R^T,R^L,S$ and also $P$ 
in Eq.(\ref{eq_p}), $3)$~solve Eq. (\ref{eq_proff}) and (\ref{eq_proft})
to obtain new profile functions, and $4)$~repeat $1)-3)$ until the 
self-consistency is attained. 

\begin{figure}
\includegraphics[height=6.5cm, width=9.5cm]{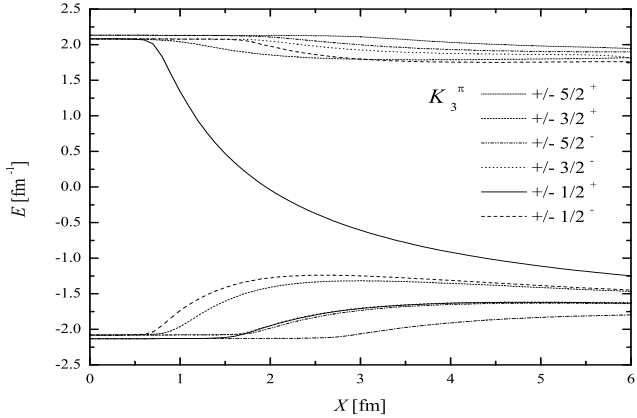}
\caption{\label{fig:Sf2} Spectrum of the quark orbits 
with $B=2$ as a function of the soliton size parameter $X$.}
\end{figure}

\begin{figure}
\includegraphics[height=6.5cm, width=9.5cm]{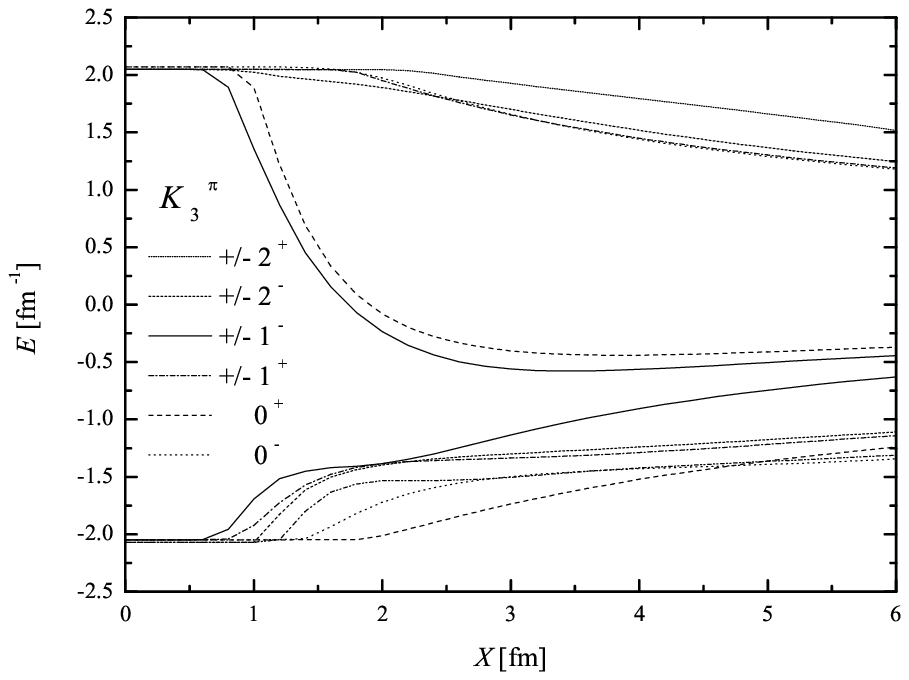}
\caption{\label{fig:Sf3} Spectrum of the quark orbits 
with $B=3$.}
\end{figure}

\begin{figure}
\includegraphics[height=6.5cm, width=9.5cm]{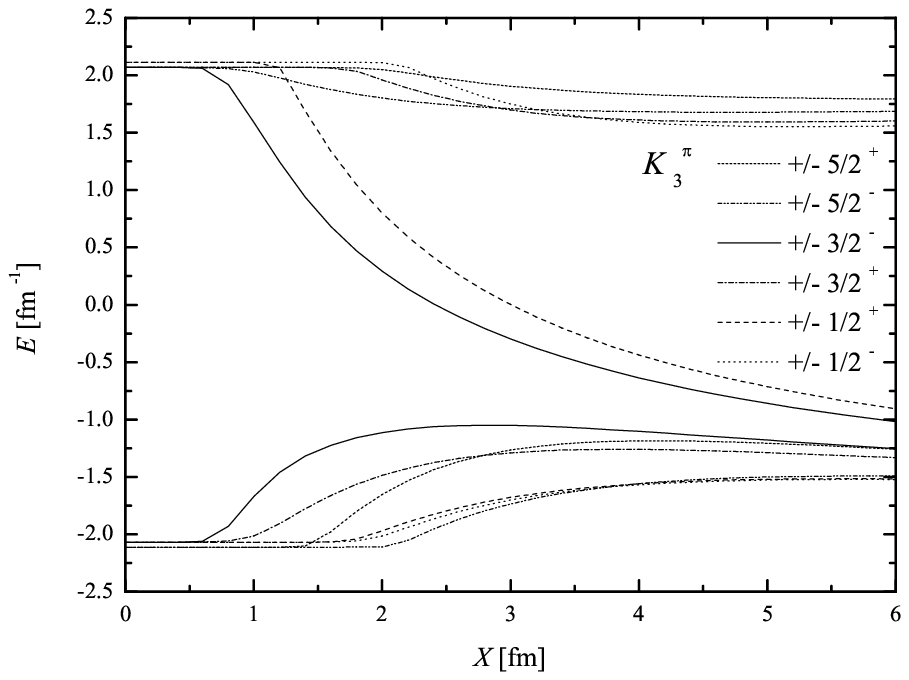}
\caption{\label{fig:Sf4} Spectrum of the quark orbits 
with $B=4$.}
\end{figure}

\begin{figure}
\includegraphics[height=6.5cm, width=9.5cm]{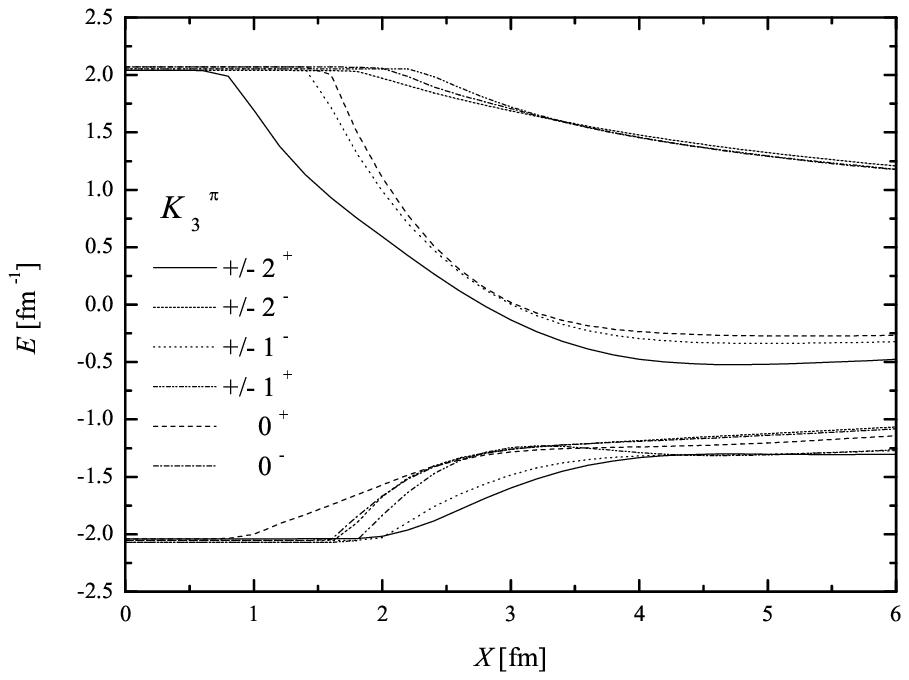}
\caption{\label{fig:Sf5} Spectrum of the quark orbits 
with $B=5$.}
\end{figure}

In Figs. \ref{fig:Sf2}-\ref{fig:Sf5}, we show the spectrum of the 
quark orbits in the background of chiral fields with winding number 
$m_{{\rm w}}=2-5$, as a function of the size parameter $X$.
The profile functions are parameterized by $X$ as

\begin{eqnarray}
	F(\rho,z)&=&-\pi+\pi\sqrt{\rho^2+z^2}/X~~{\rm for}~~X\le\sqrt{\rho^2+z^2} 
	\nonumber \\
	&=&0\hspace{3.1cm}{\rm otherwise}, \\
	\Theta(\rho,z)&=&\tan^{-1}(\rho/z).
\end{eqnarray}  
To examine the spectrum in detail, let us consider the Hamiltonian 
defined in (\ref{hamiltonian}).
For the axially symmetric chiral field in Eq.(\ref{ansatz}), this 
Hamiltonian commutes with the third component of the grand spin 
operator $K_{3}$ and the time-reversal operator $T$. 
These are specifically, 
\begin{eqnarray}
 	 &&K_3=L_3+\frac{1}{2} \sigma_3 +\frac{1}{2} m_{{\rm w}}\tau_3 \\
	 &&T=i \gamma_1 \gamma_3 \cdot i \tau_1 \tau_3 C
\end{eqnarray}
where $L_3$, $\sigma_3$, and $\tau_3$ are respectively 
the third component of orbital angular momentum, spin, and isospin operator,  
and $C$ is a charge conjugation operator. 
The parity operator is defined by $P=\gamma_0$ for odd $B$, 
and $P=\gamma_0\tau_3$ for even $B$.
Thus, the eigenvalues of the Hamiltonian can be specified 
by the magnitude of $K_3$ and the parity $\pi=\pm$. 
We have $K_3 = 0, \pm1, \pm2, \pm3,\cdots $ for odd $B$, 
and $K_{3}=\pm\frac{1}{2}, \pm\frac{3}{2}, 
\pm\frac{5}{2}, \pm\frac{7}{2}, \cdots $ for even $B$.  
Since the Hamiltonian 
is invariant under time reverse, the states of 
$+K_3$ and $-K_3$ are degenerate in energy. 

Fig.\ref{fig:Sf2} shows the quark spectrum with $m_{{\rm w}}=2$. 
The bound states diving into negative region are doubly 
degenerate with $K_3=\pm\frac{1}{2}$. 
Thus putting $N_c=3$ valence quarks on each of the bound levels, we have
the $B=2$ soliton solution. 
For $m_{{\rm w}}=3$, the spectrum of $K_3 = \pm 1^{-}$(double degeneracy) and
 $K_3 = 0^{+} $ states dive into negative-energy region. 
Thus, we have the $B=3$ soliton solution. 
For $m_{{\rm w}}=4$, the spectrum of $K_3 = \pm {\frac{1}{2}}^{+}$ 
and $K_3 = \pm {\frac{3}{2}}^{-}$(both doubly degenerate) states 
dive into negative region. Thus, we have the $B=4$ soliton solution.
For $m_{{\rm w}}=5$, the spectrum of $K_3 = \pm 2^{+}$(double), 
$K_3 = \pm 1^{-}$(double) and $K_3 = 0^{+} $ states dive into 
negative-energy region. Thus we have the $B=5$ soliton solution. 
These results confirm that the baryon number of the soliton is identified 
with the number of diving levels occupied by $N_c$ valence quarks.
It can be seen that the degeneracy which occurs due to symmetry of 
the chiral field reduces the number of states and hence makes large shell gaps. 
This observation indicates that degeneracy in spectrum contribute to 
make classical energies of the soliton solutions lower. 
In fact, our $B=2$ solution which is considered to be the minimum energy 
soliton from the study of the $B=2$ skyrmions provides the maximum degeneracy 
in spectrum. It will be interesting to study minimum solutions from this point 
of view. 

\begin{figure*}
\includegraphics[height=10cm, width=14cm]{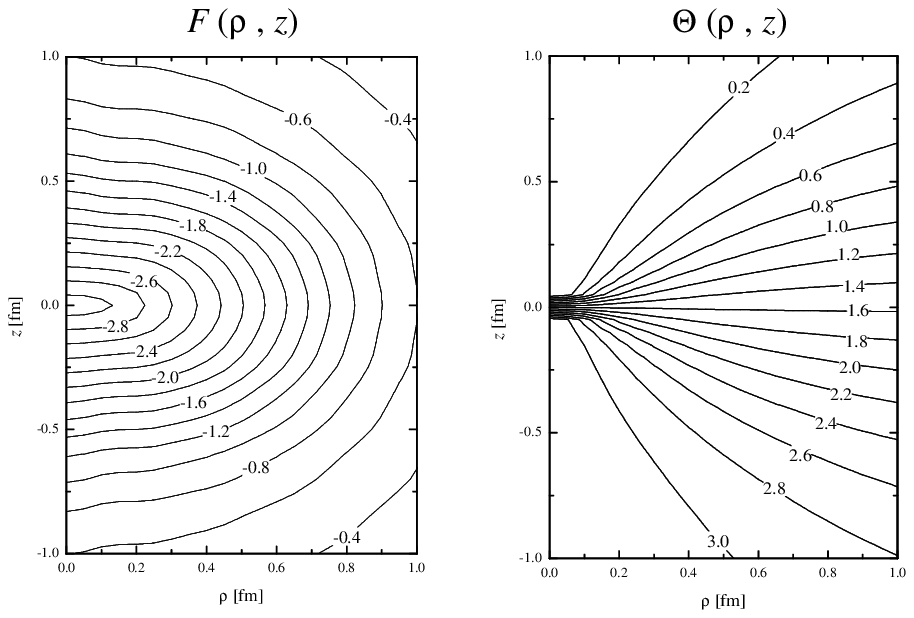}
\caption{\label{fig:Pf2} Contour plot of the profile functions
with $B=2$.}
\end{figure*}
\begin{figure*}
\includegraphics[height=10cm, width=14cm]{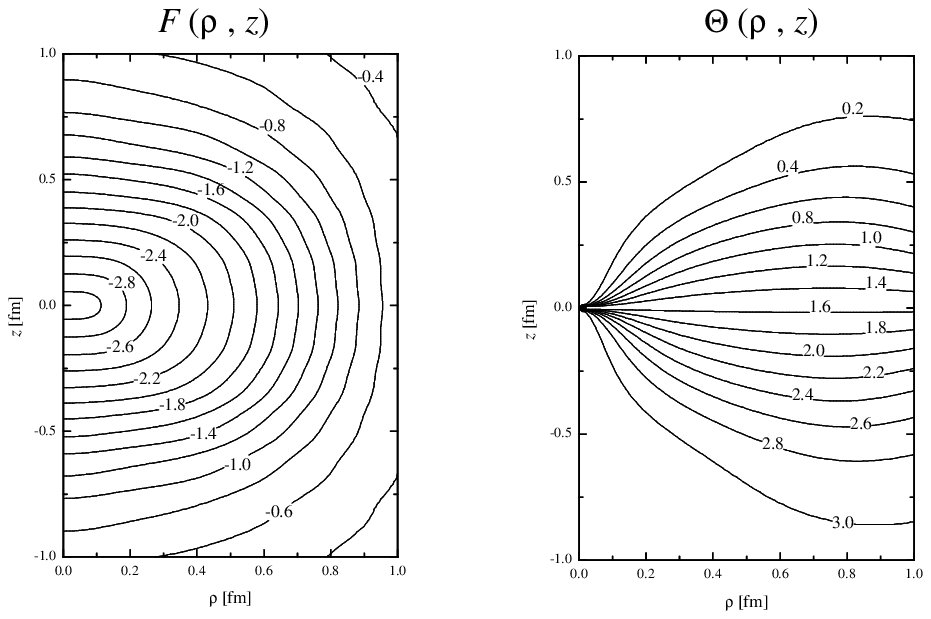}
\caption{\label{fig:Pf3} Contour plot of the profile functions
with $B=3$.}
\end{figure*}
\begin{figure*}
\includegraphics[height=10cm, width=14cm]{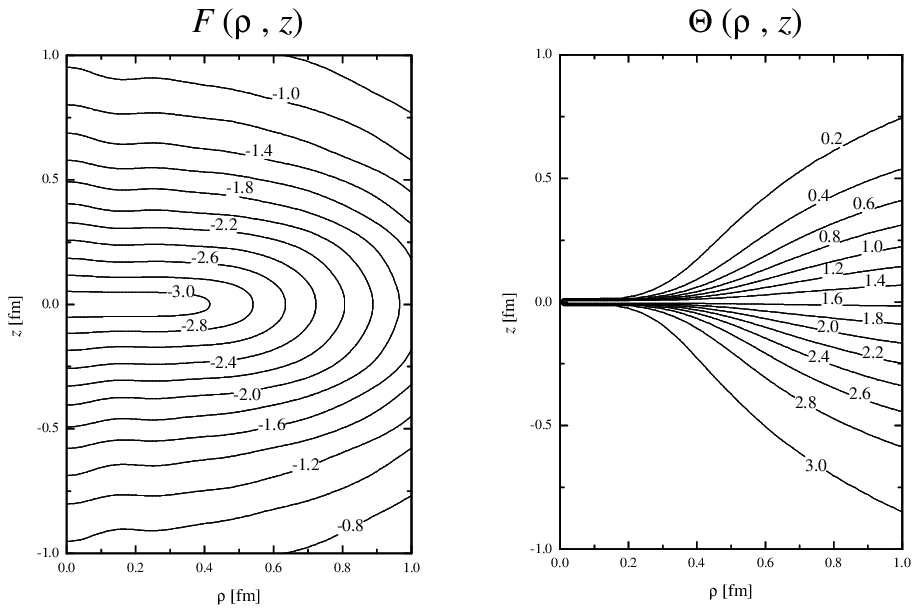}
\caption{\label{fig:Pf4} Contour plot of the profile functions
with $B=4$.}
\end{figure*}
\begin{figure*}
\includegraphics[height=10cm, width=14cm]{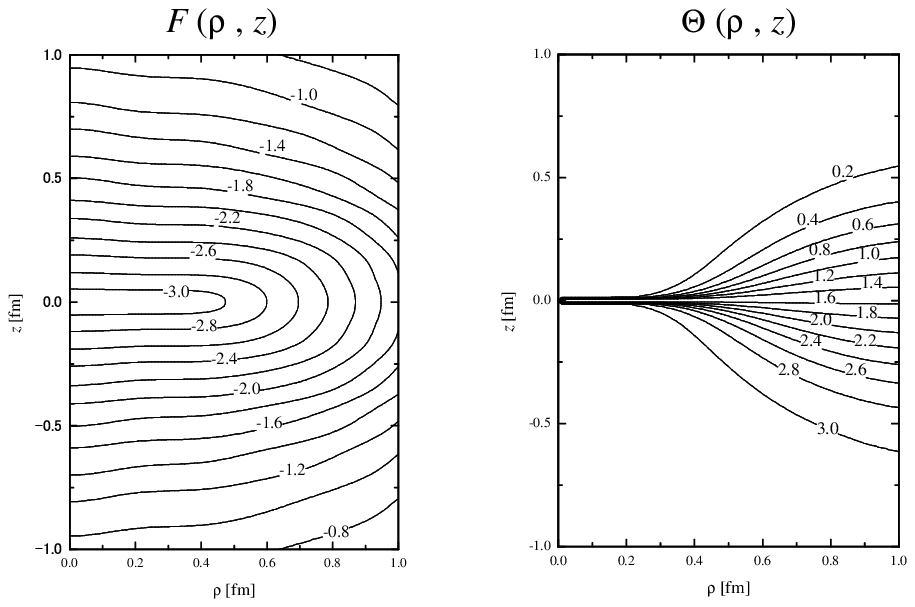}
\caption{\label{fig:Pf5} Contour plot of the profile functions
with $B=5$.}
\end{figure*}

\begin{table}
\caption{\label{tab:claspectra}The classical mass for 
$B=1\sim 5$ ($[MeV]$).}
\begin{ruledtabular}
\begin{tabular}{cccccccc}
$B$ & \multicolumn{5}{c}{Valence} & Vacuum & Total \\ \hline 
1 & 173 &     &     &     &     & 674  & 1192 \\
2 & 173 & 173 &     &     &     & 1166 & 2204 \\
3 & 173 & 173 & 298 &     &     & 1561 & 3493 \\
4 & 106 & 106 & 232 & 232 &     & 2727 & 4753 \\
5 & 145 & 145 & 319 & 319 & 409 & 2537 & 6543 \\
\end{tabular}
\end{ruledtabular}
\end{table}
\begin{table}
\caption{\label{tab:table2}The mean radius and 
mean root square radius in each baryon number for $B=1- 5$
.(The result of $B=1$ is quoted by Ref.~\cite{wakamatsu91}).}
\begin{ruledtabular}
\begin{tabular}{ccc}
\hspace{5mm}B&$\langle \rho \rangle $$[fm]$&$\sqrt{\langle r^2 
\rangle}$[$fm$]~~~~~~~\\
\hline
\hspace{5mm}1 &       & 0.785 \\
\hspace{5mm}2 & 0.672 & 0.821 \\
\hspace{5mm}3 & 0.659 & 0.854 \\
\hspace{5mm}4 & 0.971 & 1.140 \\
\hspace{5mm}5 & 1.048 & 1.225 \\
\end{tabular}
\end{ruledtabular}
\end{table}

\begin{figure*}
\includegraphics[height=14cm, width=14cm]{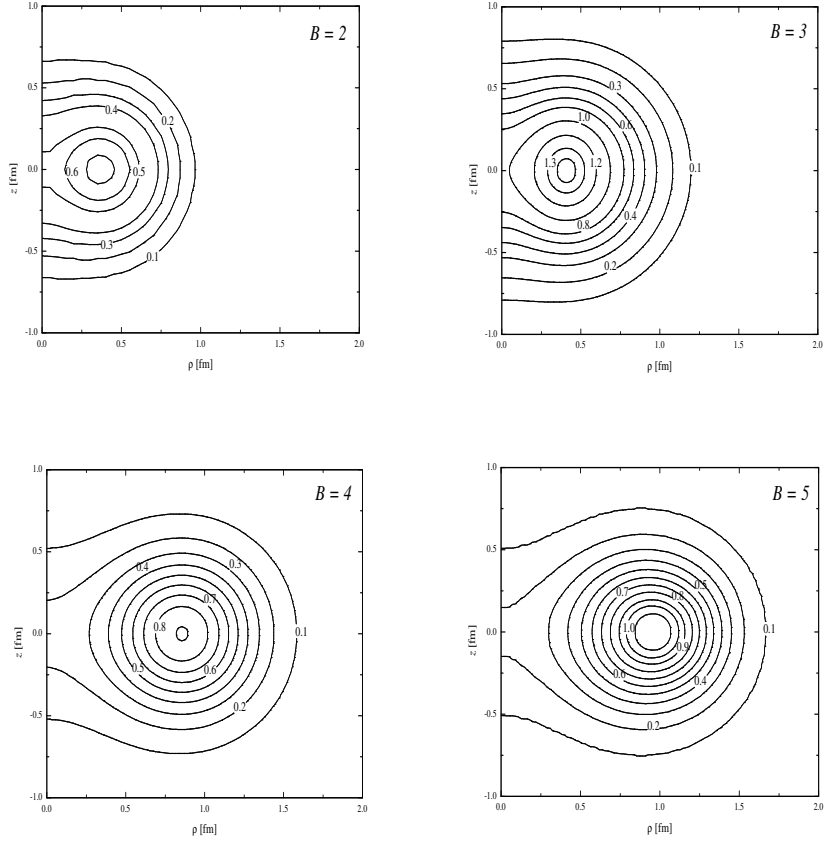}
\caption{\label{fig:Bd}baryon number density [$fm^{-3}$].}
\end{figure*}

The baryon number density is defined by the zeroth component of 
the baryon current \cite{reinhardt88}, 
\begin{equation}
	\langle B \rangle =\frac{1}{N_c}\langle \bar{\psi}\gamma_0 \psi \rangle
	={\langle B \rangle}_{\rm val}+{\langle B \rangle}_{0}
\end{equation}
where
\begin{eqnarray}
	&&\langle B \rangle_{{\rm val}}=\frac{1}{N_{c}}\sum_{i}\int d\varphi\, 
	\phi_{i}^{\dagger}(\rho,\varphi,z)\phi_{i}(\rho,\varphi,z) \nonumber\\
	&&\langle B \rangle_{0}=\frac{1}{N_{c}}\biggl[
	\sum_{\mu}{\cal N}(E_{\mu})\sgn (E_{\mu})
	\nonumber \\
	&&\hspace{1cm}\times \int d\varphi \, \phi_{\mu}^{\dagger}
	(\rho,\varphi,z)\phi_{\mu}(\rho,\varphi,z) 
	\nonumber \\
	&&\hspace{0.5cm}-\sum_{\mu}{\cal N}(E_{\mu}^{(0)})
	\sgn (E_{\mu}^{(0)})  
	\nonumber \\
	&&\hspace{1cm}\times \int d\varphi \phi_{\mu}^{(0)\dagger}(\rho,\varphi,z)
	\phi_{\mu}^{(0)}(\rho,\varphi,z) \biggr]\,.\nonumber
\end{eqnarray}
The contour plot of the baryon number density for each baryon number
is shown in Fig.~\ref{fig:Bd}. It can be seen that they have toroidal 
in shape. 

The mean radius ${\langle \rho \rangle}$ is given by 
\begin{equation}
	\langle \rho \rangle ={\langle \rho \rangle}_{\rm val}
	+{\langle \rho \rangle}_{0}.
\end{equation}
where
\begin{eqnarray*}
	&&{\langle \rho \rangle}_{\rm val} = \frac{1}{m_{{\rm w}}}
	\sum_i \int \rho d \rho d z d 
	\varphi \rho  \phi_{i}^{\dagger} (\rho,\varphi,z)
	\phi_{i}(\rho,\varphi,z)\\
	&&{\langle \rho \rangle}_{0} = \frac{1}{m_{{\rm w}}} 
	\sum_\mu {\cal N}(E_\mu) 
	\sgn(E_{\mu})  \\
	&&\hspace{1cm}\times \int \rho d \rho d z d \varphi 
	\rho \phi_{\mu}^{\dagger} 
	(\rho,\varphi,z)\phi_{\mu}(\rho,\varphi,z)\,.
\end{eqnarray*}
The root mean square radius is given by 
\begin{equation}
	\sqrt{\langle r^2 \rangle} =\sqrt{\langle r^2 \rangle_{\rm val}}+
	\sqrt{\langle r^2 \rangle_{0}}
\end{equation}
where
\begin{eqnarray*}
	&& \langle r^2 \rangle_{\rm val} = \frac{1}{m_{{\rm w}}}\sum_i \int \rho d \rho d z d 
	\varphi ({\rho}^2 + z^2) \phi_{i}^{\dagger} (\rho,\varphi,z)\phi_{i}
	(\rho,\varphi,z)  \\
	&&\langle r^2 \rangle_{0} = \frac{1}{m_{{\rm w}}} \sum_\mu {\cal N}(E_\mu) 
	\sgn(E_{\mu}) \hspace{1cm} \\
	&&~~\times \int \rho d \rho d z d \varphi ({\rho}^2 + z^2) 
	\phi_{\mu}^{\dagger} (\rho,\varphi,z)\phi_{\mu}(\rho,\varphi,z)\,.
\end{eqnarray*}
These values for each baryon number are shown in Table~\ref{tab:table2}.  

\section{\label{sec:level3}Quantization}
\subsection{Rotational Zero Mode Quantization}

The solitons that we obtained in the previous section are 
classical objects and therefore must be 
quantized to assign definite spin and isospin to them.  
Quantization of the solitons can be performed semiclassically 
for their rotational zero modes. 
For the hedgehog soliton, because of its topological structure, 
a rotation in isospin space is followed by a simultaneous spatial 
rotation. For the axially symmetric soliton, 
there are six rotational zero modes by rotations of  
iso-degrees of freedom and spatial rotations. 

Let us introduce the ``dynamically rotated'' chiral fields 
\cite{braaten88}  
\begin{equation}
	U(\bm{x},t)=A(t)U(\bm{x}')A(t)^{\dagger},~~{{x}^{i}}'  = 
	{\Xi}^{i}_j[B(t)]x^{j}
	\label{eq:35}
\end{equation}
where 
\begin{equation}
{\Xi}^{i}_j[B(t)] =\frac{1}{2}Tr[{\sigma}^{i}B(t){\sigma}_{j}B(t)^{\dagger}] \,,
\end{equation}
and $A(t)$ and $B(t)$ are time-dependent $SU(2)$ matrices generating 
an iso-rotation and a spatial rotation respectively. 
By transforming the rotating frame of reference, the Dirac operator 
with Eq.~(\ref{eq:35}) can be written as 
\begin{eqnarray}
	\tilde{iD}&=& i \gdhi - M U^{\gamma_5}({\bm x},t) \nonumber \\
	&=& A(t)S(t)^{\dagger}{\gamma}^0 [i {\partial}_t + i 
	\tilde{{\gamma}^{0}}\tilde{{\gamma}^{k}} {\partial_k}'-U^{\gamma_5}(\bm{x}') +
	iA^{\dagger} \dot{A} \nonumber\\
	&&\hspace{3cm} +iS^{\dagger} \dot{S}]S(t)A(t)^{\dagger}
	\label{eq:37}
\end{eqnarray}
where 
\begin{equation}
	\tilde{{\gamma}^{\mu}}= {\Lambda}^{\mu}_{\nu} S {\gamma}^{\nu} S^{\dagger} 
	=  \left( 
	\begin{array}{c}
	{\gamma}^{0} \\
	{\gamma}^{k} + (\bm{r}' \times \dot{\bm{\theta}})^{k}{\gamma}^{0} 
	\end{array}\right) \,,
	\label{eq:38}
\end{equation}
and $S(t)$ is the rotation operator for the Dirac field 
and $\bm{\theta}$ is an angle of the spatial rotation. 
Note that the gamma matrices $\tilde{{\gamma}^{\mu}}$ explicitly depend 
on the coordinates and do not transform as a contravariant vector
~\cite{koepf89}. 
Substituting Eq.~(\ref{eq:38}) into Eq.~(\ref{eq:37}), one obtains
\begin{eqnarray}
	\tilde{iD}= A(t)S(t)^{\dagger}{\gamma}^0 [i {\partial}_t -
	H(U^{{\gamma}_{5}}) + {\Omega}_{A} + {\Omega}_{B} ]S(t)A(t)^{\dagger} 
	\nonumber \\
\end{eqnarray}
where 
\begin{eqnarray}
	&&\hspace{1cm}\Omega_{A}=i{A^{\dagger}} \dot{A}=\frac{1}{2} \Omega^{a}_{A} \tau_a \\
	&&\Omega_{B}=i{S^{\dagger}} \dot{S} + (\bm{r} \times \bm{p}) \cdot 
	\dot{\bm{\theta}}= \Omega^{a}_{B} J_b
\end{eqnarray}
with $J_{a}=1/2 {\epsilon}_{abc} {\gamma}^{b} {\gamma}^{c} - i( \bm{r} 
\times \nabla )_{a}$. 
$\Omega_{A}$ and $\Omega_{B}$ are the angular velocity operators for an isorotation 
and for a spatial rotation respectively.   
Assuming that the rotation of the soliton is adiabatic,  
we shall expand the effective action $S_{eff}$ around the classical 
solution $U_{S}$ with respect to the angular momentum velocity 
${\Omega}_{A}$ and ${\Omega}_{B}$ up to second order~\cite{biedenharn85} 
\begin{eqnarray}
	&&S_{\rm eff}(U) \nonumber \\
	&&= S_{\rm eff}(U_S)-iN_c\Sp\left[
	\log \bigl(i {\partial}_t-H(U^{{\gamma}_{5}}_{S}) 
	+ {\Omega}_{A} + {\Omega}_{B}\bigr)\right] \nonumber \\
	&&\;\;\;-\Sp\left[\log (i {\partial}_t 
	-H(U^{{\gamma}_{5}}_{S})\right]
\end{eqnarray}
With the proper-time regularization, we have 
\begin{eqnarray}
	&&S^{\rm reg}_{\rm eff}(U)= S^{\rm reg}_{\rm eff}(U_S) \nonumber \\
	&&+\frac{1}{2} \int dt \bigl[I^{AA}_{0,ab} {\Omega}_{A}^{a}(t) {\Omega}_{A}^{b}(t) 
  	+ I^{AB}_{0,ab} {\Omega}_{A}^{a}(t) {\Omega}_{B}^{b}(t)
	\nonumber \\
	&&+ I^{BA}_{0,ab} {\Omega}_{B}^{a}(t) {\Omega}_{A}^{b}(t)
	+ I^{BB}_{0,ab} {\Omega}_{B}^{a}(t) {\Omega}_{B}^{b}(t) \bigr]
\end{eqnarray}
where $I_{0}$s are the vacuum sea contributions to the moments of inertia
defined by 
\begin{eqnarray*}
	&&I^{AA}_{0,ab} = \frac{1}{8}N_c \sum_{n,m}f(E_m,E_n,\Lambda) 
	{\langle n|{\tau}_a | m \rangle} 
	{\langle m|{\tau}_b |n \rangle} \\
	&&I^{AB}_{0,ab} = \frac{1}{4}N_c \sum_{n,m}f(E_m,E_n,\Lambda)
 	{\langle n|{\tau}_a | m \rangle} 
	{\langle m|{J}_b |n \rangle} \\
	&&I^{BA}_{0,ab} = \frac{1}{4}N_c \sum_{n,m}f(E_m,E_n,\Lambda)
 	{\langle n|{J}_a | m \rangle} 
	{\langle m|{\tau}_b |n \rangle} \\
	&&I^{BB}_{0,ab} = \frac{1}{2}N_c \sum_{n,m}f(E_m,E_n,\Lambda)
 	{\langle n|{J}_a | m \rangle} 
	{\langle m|{J}_b |n \rangle}  
	\label{eq:inertiao}
\end{eqnarray*}
with the cutoff function $f(E_m,E_n,\Lambda)$ 
\begin{eqnarray}
	&&f(E_m,E_n,\Lambda) \nonumber \\
	&&~~=-\frac{2\Lambda}{\sqrt{\pi}}
	\frac{e^{-E_m^2/\Lambda^2}-e^{-E_n^2/\Lambda^2}}{E_m^2-E_n^2} \nonumber \\
	&&~~+\frac{{\rm sgn}(E_m){\rm erfc}(|E_m|/ \Lambda)
	-{\rm sgn}(E_n){\rm erfc}(|E_n|/ \Lambda)}{E_m-E_n}\,. \nonumber \\
\end{eqnarray}
Similarly, for the valence quark contribution to the moments of inertia, 
we have
\begin{eqnarray}
	&&I^{AA}_{{\rm val},ab} = \frac{1}{2}N_c \sum_{m \neq \val} 
	\frac{{\langle \val|{\tau}_a | m \rangle} {\langle m|{\tau}_b |\val \rangle}}
	{E_m - E_{\rm val}} \nonumber \\
	&&I^{AB}_{{\rm val},ab} = N_c \sum_{m \neq \val} 
	\frac{{\langle \val|{\tau}_a | m \rangle}{\langle m|{J}_b | \val \rangle}}
	{E_m - E_{\rm val}} \nonumber \\
	&&I^{BA}_{{\rm val},ab} = N_c \sum_{m \neq \val} 
	\frac{{\langle \val|{J}_a | m \rangle} {\langle m|{\tau}_b | \val \rangle}}
	{E_m - E_{\rm val}} \nonumber \\
	&&I^{BB}_{{\rm val},ab} = 2N_c \sum_{m \neq \val} 
	\frac{{\langle \val|{J}_a | m \rangle} {\langle m|{J}_b | \val \rangle}}
	{E_m - E_{\rm val}}.
	\label{eq:inertiav}
\end{eqnarray}
The total moments of inertia are then given by the sum of the vacuum 
and valence as
\begin{eqnarray*}
	&&I^{AA}_{ab} = I^{AA}_{{\rm val},ab} + I^{AA}_{0,ab} \\
	&&I^{AB}_{ab} = I^{AB}_{{\rm val},ab} + I^{AB}_{0,ab} \\
	&&I^{BA}_{ab} = I^{BA}_{{\rm val},ab} + I^{BA}_{0,ab} \\
	&&I^{BB}_{ab} = I^{BB}_{{\rm val},ab} + I^{BB}_{0,ab}.
\end{eqnarray*}
From axial symmetry of the system, following relations are derived 
\begin{eqnarray}
	&&\hspace{1.8cm}I_{ij}= 0, ~~i\neq j, \nonumber\\
	&&\hspace{1cm}I_{11}^{AA} =  I_{22}^{AA}, ~~ I_{11}^{BB} =  I_{22}^{BB}, \nonumber\\
	&&\hspace{5mm}I_{11}^{AB} =  I_{22}^{AB} = I_{11}^{BA} =  I_{22}^{BA} = 0 , \nonumber\\
	&& I_{33}^{BB} = m^{2} I_{33}^{AA}, ~ I_{33}^{AB} =  I_{33}^{BA} = 
	-m_{{\rm w}} I_{33}^{AA}.
	\label{eq:50}
\end{eqnarray}
Theoretically, these moments of inertia can be computed using the 
eigenstates of Eq.(\ref{eigenequation}). However, due to the difference 
of the boundary conditions between the initial and final states of 
the matrix element, the moments of inertia acquire nonzero values 
with vanishing pion fields. 
To overcome this problem, we make the following replacement~\cite{goeke91}
\begin{eqnarray}
	{\langle n |{J}_a | m \rangle} &\rightarrow& {\langle n |
	[H(U^{\gamma_5}_S),{J}_a] | m \rangle}/(E_n - E_m) \nonumber\\
	&=&{\langle n |[MU^{\gamma_5}_S ,{l}_a] | m \rangle}/(E_n - E_m)
	\label{eq:replacement}
\end{eqnarray}
where ${l}_a=-i(\bm{r} \times \nabla)_a$.  
Unless the Hamiltonian explicitly depend on the coordinates, 
the numerator vanishes with vanishing pion fields. 
The spurious contributions to the moment of 
inertia can be removed in this way.

The quantization conditions for the collective coordinates, $A(t)$ and $B(t)$, 
define a body-fixed isospin operator ${\bf K}$ and a body-fixed angular 
momentum operator ${\bf L}$ as 
\begin{eqnarray}
	&&I^{AA}_{ab} {\Omega}^{b}_{A} + I^{AB}_{ab} {\Omega}^{b}_{B} \rightarrow -
	\tr \bigg( A \frac{ {\tau}_a }{2} \frac{\partial}{\partial A}  \biggr) 
	\equiv -K_a \\
	&&I^{BA}_{ab} {\Omega}^{b}_{A} + I^{BB}_{ab} {\Omega}^{b}_{B} \rightarrow 
	\tr \bigg( \frac{ {\sigma}_a }{2} B \frac{\partial}{\partial B}  \biggr) 
	\equiv -L_a.
	\label{eq:48}
\end{eqnarray}
These are related to the usual coordinate-fixed isospin 
operator $I_a$ and coordinate-fixed angular momentum $J_a$ operator by 
transformations,
\begin{equation}
I_{a}=- \Xi^{b}_{a}[A(t)]K_b,~~~J_{a}=- \Xi^{b}_{a}[B(t)]^{T}L_b .
\label{eq:49}
\end{equation}    
   
To estimate the quantum energy corrections, let us introduce 
the basis functions of the spin and isospin operators which
were inspired from the cranking method for 
nuclei~\cite{bohr,weigel86},
\begin{eqnarray*}
	&&\langle A,B|{i{i}_{3}{k}_{3},j{j}_{3}{l}_{3}}
	\rangle  \\
	&&\hspace{1cm}=\frac{\sqrt{(2i+1)(2j+1)}}{8{\pi}^{2}}
	D^{i~~\ast}_{i_{3}k_{3}}(A)D^{j~~\ast}_{j_{3},-m_{\rm w}k_{3}}(B) 
\end{eqnarray*}  
where $D$ is the Wigner rotation matrix. Then, we find the 
quantized energies of the soliton as 
\begin{eqnarray}
	E&=&E_{static}+\frac{1}{2I_{11}^{AA}}i(i+1)
	+\frac{1}{2I_{11}^{BB}}j(j+1) \nonumber \\ 
	&& +\frac{1}{2}\left(\frac{1}{I_{33}^{AA}}-\frac{1}{I_{11}^{AA}}
	-\frac{m^2_{\rm w}}{I_{11}^{BB}}\right)k_{3}^{2} \label{eq:51}
\end{eqnarray}
where $i(i+1), j(j+1)$ and $k_3$ are eigenvalues of the Casimir operators 
$\bm{I}^{2}$ and $\bm{J}^{2}$, and the operator $\bm{K}_{3}$, respectively.

On Table~\ref{tab:table3} are the results of our calculation 
of moments of inertia, $I_{11}^{AA}, I_{11}^{BB}$ and $I_{33}^{AA}$, 
with $B=2-5$. It is instructive to compare our results
with the Skyrme model~\cite{braaten88} where $U_{11}=0.104, 
V_{11}=0.163$ and $U_{33}=0.0709$ which are correspondingly 
our $I_{11}^{AA}, I_{11}^{BB}$ and $I_{33}^{AA}$. 
They are qualitatively in good agreement.  

\begin{table}
\caption{\label{tab:table3}Moments of inertia.}
\begin{ruledtabular}
\begin{tabular}{ccccc}
~~~B& & Valence & Sea & Total~~~ \\
\hline
~~~2&$I^{AA}_{11}$&0.00773& 0.00363 &0.01136~~~\\
~~~ &$I^{BB}_{11}$&0.01141& 0.00464 &0.01605~~~\\
~~~ &$I^{AA}_{33}$&0.00429& 0.00125 &~0.00554~~~ \\
\hline
~~~3&$I^{AA}_{11}$&0.01231& 0.00280 &0.01511~~~\\
~~~ &$I^{BB}_{11}$&0.02174& 0.00384 &0.02558~~~\\
~~~ &$I^{AA}_{33}$&0.00594& 0.00027 &~0.00622~~~ \\
\hline
~~~4&$I^{AA}_{11}$&0.01408& 0.00959 &0.02366~~~\\
~~~ &$I^{BB}_{11}$&0.04272& 0.01245 &0.05517~~~\\
~~~ &$I^{AA}_{33}$&0.01172& 0.00074 &~0.01246~~~ \\
\hline
~~~5&$I^{AA}_{11}$&0.02786& 0.00716 &0.03502~~~\\
~~~ &$I^{BB}_{11}$&0.12124& 0.01112 &0.13236~~~\\
~~~ &$I^{AA}_{33}$&0.01368& 0.00007 &~0.01375~~~ \\
\end{tabular}
\end{ruledtabular} \end{table}

\subsection{Finkelstein-Rubinstein constraints}
If a multi-skyrmion describes atomic nuclei upon quantization, 

it has to be quantized as a boson or as a fermion whether $B$ is even or odd. 
This requirement is implemented in the form of Finkelstein-Rubinstein 
(FR) constraints \cite{fr}. The FR constraints for the rational map ansatz 
was constructed in \cite{irwin} and \cite{krusch} and applied to predict 
the ground states of skyrmions up to $B=22$. In this section, we shall 
apply the FR constraints for the rational map ansatz directly to our axially 
symmetric multi-skyrmions and obtain their ground states. 
 
Following the notation in \cite{krusch}, let $g$ be a rotation by $\alpha$ 
around ${\bf n}$ followed by an isorotation by $\beta$ around ${\bf N}$. 
Then the FR constraints can be defined as 
\begin{eqnarray}
	\exp(-i\alpha {\bf n}\cdot{\bf J})\exp(-i\beta {\bf N}\cdot{\bf I})
	\psi =  \chi_{FR}(g)\psi \label{}
\end{eqnarray}
where 
$$ \chi_{FR}(g)=\begin{cases} 1 & \text{if contractible}\\
-1 & \text{otherwise.} \end{cases}$$ 
and, ${\bf J}$ and ${\bf I}$ are space-fixed spin and isospin operators respectively. 
$\psi$ is the wave function which transforms under a tensor product 
of rotations and isorotations. In particular, a closed loop is noncontractible 
for odd $B$ and contractible for even $B$, which is consistent with spin statistics. 
Consequently, quantum numbers $I$ and $J$ are half-integers for odd $B$ and 
integers for even $B$. 

In order to construct the ground states for a given baryon number $B$, 
let us define $N(L(\alpha, \beta))$ as a homotopy invariant for a loop $L$
generated by rotations by $\alpha$ and isorotations by $\beta$. 
Then, for the axially symmetric rational map of degree $B$, it is given by \cite{krusch}
\begin{eqnarray}
	N(L(\alpha, \beta))=\frac{B}{2\pi}(B\alpha-\beta) \,.\label{n(l)}
\end{eqnarray}
It can be shown that $N$ (mod $2$) determines if the loop is contractible or 
not in the same sense as $B$ (mod $2$). Therefore, $N$ (mod $2$) gives the FR 
constraints for each generator of the symmetry group of the rational map. 

The axially symmetric rational map with degree $B$ is given by 
\begin{eqnarray}
	R(z)=\frac{1}{z^{B}} \, . \label{}
\end{eqnarray}
There are two symmetric generators for this rational map. 
One is a rotation by $\alpha$ followed by an isorotation by 
$\beta = B\alpha$. Substituting it into (\ref{n(l)}), one obtains $N(L(\alpha, 
B\alpha))=0$. The FR constraints for this loop is thus given by 
\begin{eqnarray}
	e^{-i\pi(L_{3}-BK_{3})}\psi = \psi \, . \label{j3}
\end{eqnarray} 
where we introduced the body-fixed spin $({\bf L})$ and isospin $({\bf K})$  
operators related to the space-fixed operators by 
orthogonal transformations. The other symmetry is $C_{2}$ with transformation 
\begin{eqnarray}
	z \rightarrow \frac{1}{z}\;, \;\;\; R(z) \rightarrow \frac{1}{R(z)}\,. \label{}
\end{eqnarray}
This corresponds to $\alpha=\beta=\pi$ and hence $N(L(\pi,\pi))=B(B-1)/2$. 
The FR constraints for this loop is 
\begin{eqnarray}
	e^{-i\pi(L_{1}+K_{1})}\psi = (-1)^{B(B-1)/2}\psi \,. \label{j1}
\end{eqnarray}

In the following we construct the ground states consistent with the derived 
FR constraints (\ref{j3}) and (\ref{j1}) for $B=2-5$ with axial symmetry. 
\begin{itemize}
	\item $B=2$ \\
	We find the FR constraints 
\begin{eqnarray}
	&& e^{-i\pi(L_{3}-2K_{3})}\psi = \psi \\ 
	&& e^{-i\pi(L_{1}+K_{1})}\psi = -\psi \,. \label{}
\end{eqnarray} 
This gives the ground state as 
$\left|J,L_{3}\right>\left|I,K_{3}\right>=\left|1,0\right>\left|0,0\right>$. \\
	\item $B=3$ \\
	We find the FR constraints 
\begin{eqnarray}
	&& e^{-i\pi(L_{3}-3K_{3})}\psi = \psi \\ 
	&& e^{-i\pi(L_{1}+K_{1})}\psi = -\psi \,. \label{}
\end{eqnarray} 
This gives the ground state as 
$\left|\frac{5}{2},\frac{3}{2}\right>\left|\frac{1}{2},\frac{1}{2}\right>$. \\ 
	\item $B=4$ \\
	We find the FR constraints 
\begin{eqnarray}
	&& e^{-i\pi(L_{3}-4K_{3})}\psi = \psi \\ 
	&& e^{-i\pi(L_{1}+K_{1})}\psi = \psi \,. \label{}
\end{eqnarray} 
This gives the ground state as $\left|0,0\right>\left|0,0\right>$. \\ 
	\item $B=5$ \\
	We find the FR constraints 
\begin{eqnarray}
	&& e^{-i\pi(L_{3}-5K_{3})}\psi = \psi \\ 
	&& e^{-i\pi(L_{1}+K_{1})}\psi = \psi \,. 
	\label{}
\end{eqnarray} 
This gives the ground state as $\left|J\right>\left|I\right>=\left|
\frac{7}{2},\frac{5}{2}\right>\left|\frac{1}{2},\frac{1}{2}\right>$. \\ 
	
\end{itemize}
Thus, for even $B$, the axially symmetric solitons are possible 
candidates of the ground states of $B$ atomic nuclei as is 
the case of the deuteron and $^{4}{He}$ while for odd $B$ 
they emerge only as excited states. 

\section{\label{sec:level4}NUMERICAL ANALYSIS}
\subsection{Eigen Equations}
In this subsection, we show the numerical analysis 
of the eigen equations in detail.  
To solve the eigenequation of the form,  
\begin{eqnarray}
	&&[-i \bm{\alpha\cdot}\nabla + \beta M (\cos F(\rho,z)  + i {\gamma}_{5} 
	\bm{\tau} \bm\cdot \hat{\bm{n}} \sin F(\rho,z)]\phi_\mu(\bm{x})\nonumber \\
	&&\hspace{5.5cm}=E_\mu\phi_\mu(\bm{x})\,,
	\label{eigenequation2}
\end{eqnarray}
we introduce the deformed harmonic oscillator spinor basis which was 
originally constructed by Gambhir {\it et al.} in the relativistic 
mean field theory for deformed nuclei~\cite{gambhir}. 
The upper and lower components of the Dirac spinors are 
expanded separately by the basis as 
\begin{equation}
	{\phi}_{\mu} (\bm{x})= 
	\left( 
	\begin{array}{c}
	f_{\mu} (\bm{x}) \\
	ig_{\mu} (\bm{x}) 
	\end{array}\right)\; 
	 =  \left( 
	\begin{array}{c}
	\sum_{a}f_{\mu a} \bm{\Phi} _{a} (\bm{x},s) \\
	i \sum_{ \tilde{a}}g_{\mu \tilde{a}} \bm{\Phi} _{\tilde{a}} (\bm{x},s)  
	\end{array}\right)\; 
	{\chi}^I_{m_{\tau}}
\end{equation}
where $\bm{\Phi} _{a} (\bm{x},s,\tau) $ are the eigefunctions of 
a deformed harmonic oscillator potential
\begin{eqnarray}
	V_{osc}(\rho,z)=\frac{1}{2}{\cal M}\omega^2_\rho \rho^2
	+\frac{1}{2}{\cal M}\omega^2_z z^2 \,,
\end{eqnarray}
and defined by   
\begin{equation}
	\bm{\Phi} _{a} (\bm{x},m_s) = \frac{1}{\sqrt{2 \pi}} \phi^{|\omega|}_{n_{r}}({\rho}) 
	{\phi}_{n_{z}}(z) e^{i\omega \varphi} {\chi}^S_{m_{s}}
\end{equation}
with
\begin{eqnarray*}
	&&{\phi}^{|\omega|}_{n_{r}}({\rho}) = N_{n_r}^{|\omega|}
	(\sqrt{{\alpha}_{\rho} \rho})^{|\omega|} e^{- \frac{1}{2} 
	{\alpha}_{\rho} {\rho}^{2}}L^{|\omega|}_{n_r}(\alpha_\rho \rho^2) \\
	&& n_{r}=0,1,2, \cdot \cdot \cdot, N_{rmax} \\
	&&{\phi}_{n_z}(z) = N_{n_z}
	 e^{- \frac{1}{2} \alpha_z z^2}H_{n_z}
	(\sqrt{{\alpha}_{z}} {z}) \\
	&& n_{z}=1,3, \cdot \cdot \cdot, 2N_{zmax}+1~~{\rm or}~~0,2,\cdots 
	, 2N_{zmax} \,,
\end{eqnarray*}
and 
\begin{equation}
	{\chi}_{+}=\left( 
	\begin{array}{c}
	1  \\
	0
	\end{array}\right)~,~~~~~~{\chi}_{-}=\left( 
	\begin{array}{c}
	0  \\
	1
	\end{array}\right) 
\end{equation} 
depending on if the eigenvalues of the third components 
of the spin $m_{s}$ (isospin $m_{\tau}$) takes $+1$ or $-1$. 
The functions, $L^{|m|}_{n_{r}}$ and 
$H_{n_{z}}$, are the associated Laguerre polynomials 
and the Hermite polynomials with the normalization constants
\begin{eqnarray}
	N_{n_r}^{|\omega|}=\sqrt{\frac{2\alpha_\rho n_r!}{(n_r+|\omega|)!}}\,\,,\,\,
	N_{n_z}=\frac{1}{\sqrt{2^{n_z}n_z!\sqrt{\frac{\pi}{\alpha_z}}}}\,.
\end{eqnarray}
These polynomials can be calculated by following recursion relations
\begin{eqnarray}
	x\frac{d}{dx}L^{\alpha}_{n}(x)=nL^{m}_{n}(x)-(n+m)L^{\alpha}_
	{n-1}(x) \\
	L^{m-1}_{n}(x)=L^{m}_{n}(x)-L^{m}_{n-1}(x)
\end{eqnarray}
and
\begin{eqnarray}
	H_{n+1}(x)-2xH_{n}(x)+2n_zH_{n-1}(x)=0\\
	\frac{d}{dx}H_{n}(x)=2nH_{n-1}(x) 
\end{eqnarray}
where constants ${\alpha}_{\rho}$ and ${\alpha}_{z}$ can be expressed
by the oscillator frequencies as
\begin{eqnarray}
\alpha_\rho=\frac{{\cal M}\omega_\rho}{\hbar}\,,\,\, 
\alpha_z=\frac{{\cal M}\omega_z}{\hbar}
\end{eqnarray}
which are free parameters chosen optimally. 
The $N_{rmax}$ and $N_{zmax}$ 
are increased until convergence is attained. 
The parity transformation rule of ${\Phi}_{\alpha}$ is given by 
\begin{equation}
	{\Phi}_{\alpha}(\rho,\varphi + \pi ,-z;s,t)={(-1)}^{\omega+{n}_{z}}{\Phi}_{\alpha}
	(\rho,\varphi,z;s,t) 
\end{equation}
where 
\begin{equation}
	H_{n_z} (-\sqrt{\alpha_z} z)={(-1)}^{n_z}{H}_{n_z} 
	(\sqrt{\alpha_z} z),
	\label{eq:61}
\end{equation}
has been used.
The parity is $+$ for $\omega+{n}_{z}=$ odd, and  
$-$ for $\omega+{n}_{z}=$ even. 

There are two sets of the complete basis for each parity. 
One is the natural basis with ${K}_{3}^{P}= {0}^{+}, {1}^{-}, {2}^{+}
, \cdots $, for odd $B$ and $K_{3}^{P} =
{\frac{1}{2}}^{+}, {\frac{3}{2}}^{-}, 
{\frac{5}{2}}^{+},\cdots $ for even $B$. 
Another is the unnatural basis with  ${K}_{3}^{P}= {0}^{-}, 
{1}^{+}, {2}^{-}, \cdots $, for odd $B$
and $K_{3}^{P} ={\frac{1}{2}}^{-}, {\frac{3}{2}}^{+}, 
{\frac{5}{2}}^{-}, \cdots $ for even $B$
The natural basis is given by 
\begin{widetext}
\begin{eqnarray}
	&&{\phi}^{\it (n)}_\mu(\bm{x})= \nonumber \\
	&&\left( 
	\begin{array}{c}
	 \displaystyle{\sum_{\alpha(0)}}f_{\alpha(0),\mu}
	 {\bm \Phi}_{\alpha(0)}(\bm{x},\uparrow_S)
	 +\displaystyle{\sum_{\alpha(1)}}f_{\alpha(1),\mu}
	 {\bm \Phi}_{\alpha(1)}(\bm{x},\downarrow_S) \\
	i \displaystyle{\sum_{\beta(0)}}g_{\beta(0),\mu}
	  {\bm \Phi}_{\beta(0)}(\bm{x},\uparrow_S)
	+i\displaystyle{\sum_{\beta(1)}}g_{\beta(1),\mu}
	  {\bm \Phi}_{\beta(1)}(\bm{x},\downarrow_S)
	\end{array}\right)\; {\chi}^I_{u}
	+\left(
	\begin{array}{c}
	 \displaystyle{\sum_{\alpha(2)}}f_{\alpha(2),\mu}
	 {\bm \Phi}_{\alpha(2)}(\bm{x},\uparrow_S)
	+\displaystyle{\sum_{\alpha(3)}}f_{\alpha(3),\mu} 
	 {\bm \Phi}_{\alpha(3)}(\bm{x},\downarrow_S)
	\\
	 i\displaystyle{\sum_{\beta(2)}}g_{\beta(2),\mu}
	  {\bm \Phi}_{\beta(2)}(\bm{x},\uparrow_S)
	+i\displaystyle{\sum_{\beta(3)}}g_{\beta(3),\mu}
	  {\bm \Phi}_{\beta(3)}(\bm{x},\downarrow_S)
	\end{array}\right)\; {\chi}^I_{d} \nonumber \\
	\label{expansion}
\end{eqnarray}
\end{widetext}
where
\begin{eqnarray*}
	&&\alpha (0) = \{n_r, n_z: {\rm odd}, 
	~\omega_0\equiv K_3-1/2-m_{\rm w}/2\}  \\
	&&\alpha (1) = \{n_r, n_z: {\rm even},
	\omega_1\equiv K_3+1/2-m_{\rm w}/2\}  \\
	&&\alpha (2) = \{n_r, n_z: {\rm even},
	\omega_2\equiv K_3-1/2+m_{\rm w}/2\}  \\
	&&\alpha (3) = \{n_r, n_z: {\rm odd},
	~\omega_3\equiv K_3+1/2+m_{\rm w}/2\}  \\
\end{eqnarray*}
and 
\begin{eqnarray*}
	&&\beta (0) = \{n_r, n_z: {\rm even}, 
	\omega_0\equiv K_3-1/2-m_{\rm w}/2\} \\
	&&\beta (1) = \{n_r, n_z: {\rm odd},
	~\omega_1\equiv K_3+1/2-m_{\rm w}/2\} \\
	&&\beta (2) = \{n_r, n_z: {\rm odd},
	~\omega_2\equiv K_3-1/2+m_{\rm w}/2\} \\
	&&\beta (3) = \{n_r, n_z: {\rm even}, 
	\omega_3\equiv K_3+1/2+m_{\rm w}/2\}. \\
\end{eqnarray*}
The unnatural basis $\phi_{\mu}^{(u)}$ is given by replacing, 
$\alpha \leftrightarrow \beta$ in (\ref{expansion}).

\subsection{Matrix elements of the eigenequation} 
By using the natural and unnatural basis, the eigenvalue problem
in Eq.(\ref{eigenequation2}) can be reduced to a symmetric matrix  
diagonalization problem.  

Let us calculate the matrix elements of the Hamiltonian below. 
For the kinetic term 
\begin{eqnarray*}
	{\bm{\alpha \cdot p}}=\left(
	\begin{array}{cc}
	0 & \bm{\sigma\cdot p} \\
	\bm{\sigma\cdot p} & 0
	\end{array}
	\right) \,,
\end{eqnarray*}
we have 
\begin{eqnarray}
	&&\langle
	{\bm \Phi}_{\alpha(0)}|{\bm{\sigma \cdot p}}|i{\bm \Phi}_{\beta'(0)}\rangle
	\nonumber 
	\\
	&&= \frac{1}{2 \pi} \int d^{3} x   {\phi}^{|\omega_0|}_{n_r}(\rho)
 	{\phi}_{n_z}(z) e^{-i\omega_0 \varphi } 
	\nonumber \\
	&&\hspace{0.5cm}\times \Bigl(\frac{\partial}{\partial z}\Bigr) 
	{\phi}^{|\omega_0'|}_{n_r'} (\rho) 
	{\phi}_{n_z'} (z) e^{i\omega_0'\varphi}
	\nonumber
	\\
	&&
	= {\delta}_{n_r n_r'}( {N_{n_z} N_{n_z'}}
	\sqrt{\alpha_z} n_z' \frac{1}{N^{2}_{n_{z}}} 
	{\delta}_{{n_z}{n_z'-1}} \nonumber \\
	&&\hspace{0.5cm}- \frac{1}{2}{N_{n_z} N_{n_z'}}\sqrt{\alpha_z}
	\frac{1}{N^{2}_{n_{z}}} {\delta}_{{n_z} {n_z'+1}})~~\nonumber\\
	&&
	=\left\{ 
	\begin{array}{c} 
 	 \delta_{\omega_0 \omega_0'} {\delta}_{n_r n_r'} \frac{N_{n_z}'}{N_{n_z}} 
 	 \sqrt{\alpha_z} n_z' {\delta}_{{n_z}{n_z'-1}}\\
	 \delta_{\omega_0 \omega_0'} {\delta}_{{n_r} {n_r'}} (-\frac{1}{2}) \frac{N_{n_z'}}
	{N_{n_z}} \sqrt{\alpha_z}{\delta}_{n_z n_z'+1 }
	\end{array}\right.\;
	\label{eq:83}
\end{eqnarray}
and 
\begin{eqnarray}
	&&\langle
	{\bm \Phi}_{\alpha(0)}|{\bm{\sigma \cdot p}}|
	i{\bm \phi}_{\beta'(1)}\rangle
	\nonumber
	\\
	&&= \frac{1}{2 \pi} \int d^{3} x
	{\phi}^{|\omega_0|}_{n_{r}}(\rho) {\phi}_{n_{z}}(z) 
	e^{-i\omega_0\varphi }
	\nonumber \\
	&&\hspace{0.5cm}\times e^{-i \varphi}\Bigl(\frac{\partial}{\partial \rho} 
	- \frac{i}{\rho} \frac{\partial}
	{\partial \varphi}\Bigr) {\phi}^{|\omega_1'|}_{n_r'} (\rho) \phi_{n_z'} (z) 
	e^{i\omega_1'\varphi }
	\nonumber
	\\
	&&=\left\{ 
	\begin{array}{c}
	 \delta_{n_z n_z'}\sqrt{\alpha_r} (\sqrt{n_r+\omega_0+1} 
	 \delta_{n_r n_r'} +\sqrt{n_r}\delta_{n_r-1 n_r'}) \\
	(\omega_0 \ge 0 :\omega_1'=\omega_0+1>0)
	\nonumber \\
	-\delta_{n_z n_z'} \sqrt{\alpha_r}(\sqrt{n_r-\omega_0} 
	 \delta_{n_r n_r'} +\sqrt{n_r+1}
	 \delta_{n_r n_r'-1}). \\
	(\omega_0 < 0 :\omega_1'= \omega_0+1 \le 0)
	\end{array}\right.\;
	\label{eq:84} \\
\end{eqnarray} 
In the natural basis, quantum numbers $(n_z,n_z')$ 
takes values $(1,2),(3,4), \cdots$ for the upper part 
and $(1,0),(3,2), \cdots$ for the lower part. 
In the unnatural basis, $(n_{z},{n'}_{z})=(0,1),(2,3), \cdots$ 
for the upper part and $(n_{z},{n'}_{z})=(2,1),(3,2), \cdots$ 
for the lower part. 

For the potential term  
\begin{eqnarray}
	&&\beta M (\cos F(\rho,z)  + i {\gamma}_{5} 
	 \bm{\tau} \bm{\cdot} \hat{\bm{n}} \sin F(\rho,z)) \nonumber \\
	&&\hspace{0.5cm}=M\left(
	\begin{array}{cc}
	\cos F(\rho,z) & i\bm{\tau \cdot}\hat{\bm{n}}\sin F(\rho,z) \\
	-i\bm{\tau \cdot}\hat{\bm{n}}\sin F(\rho,z) & -\cos F(\rho,z) 
	\end{array}
	\right) \,, \nonumber
\end{eqnarray} 
we have 
\begin{eqnarray}
	&&\langle
	 {\bm \Phi}_{\alpha(0)}\chi^I_u| M \cos F(\rho,z)
	 |{\bm \Phi}_{\alpha'(0)}\chi^I_u \rangle 
	 \nonumber
	 \\
	&&= \int \rho d \rho dz M \cos F(\rho,z) 
	\nonumber
	\\
	&&\hspace{0.5cm}\times{\phi}^{|\omega_0|}_{n_{r}} (\rho) {\phi}_{n_{z}}(z)
	{\phi}^{|\omega_0|}_{{n'}_{r}} 
	(\rho) {\phi}_{{n'}_{z}} (z) \\
	&&
	\langle {\bm \Phi}_{\alpha(0)}\chi^I_u| M i  \bm{\tau}
	 \bm{\cdot} \hat{\bm{n}} \sin F(\rho,z)
	|i{\bm \Phi}_{\beta'(0)}\chi^I_u\rangle
	\nonumber
	\\
	&&= -\int \rho d \rho dz M \cos {\Theta}(\rho,z) \sin F(\rho,z) 
	\nonumber
	\\
	&&\hspace{0.5cm}
	\times {\phi}^{|\omega_0|}_{n_{r}} (\rho) {\phi}_{n_{z}}(z)
	  {\phi}^{|\omega_o'|}_{{n'}_{r}} 
	 (\rho) {\phi}_{{n'}_{z}} (z)
	\label{eq:92}
\end{eqnarray} 
and
\begin{eqnarray}
	&&
	\langle {\bm \Phi}_{\alpha(0)}\chi^I_u|
	M i \bm{\tau} \bm{\cdot} \hat{\bm{n}}\sin F(\rho,z)
	|i{\bm \Phi}_{\beta'(2)}\chi^I_d \rangle
	 \nonumber
	\\
	&&= -\int \rho d \rho dz M \sin {\Theta}(\rho,z) \sin F(\rho,z) 
	\nonumber
	\\
	&&\hspace{0.5cm}\times{\phi}^{|\omega_0|}_{n_{r}}(\rho)  {\phi}_{n_{z}}(z)
	  {\phi}^{|\omega_2|}_{{n'}_{r}} (\rho) {\phi}_{{n'}_{z}} (z)\,.
	\label{eq:93}
\end{eqnarray} 
Other elements can be calculated in the same manner.
In Appendix \ref{sec:levela}, we shall present these calculations of 
the matrix elements in more detail.   

\subsection{Matrix Elements for the Moments of Inertia}
To compute the matrix elements of the moments of 
inertia, we shall evaluate $\langle n| \tau_1|m \rangle , 
\langle n| \tau_3|m \rangle$ and $\langle n| J_1|m \rangle$. 
For $\langle n| \tau_1|m \rangle$, only following elements survive
\begin{eqnarray}
	&&\langle {\bm \Phi}^{\it (n)}_{\alpha (0)}\chi^I_u
	|\tau_1|{\bm \Phi}^{\it (u)}_{\alpha' (2)}\chi^I_d \rangle 
	= \langle {\bm \Phi}^{\it (n)}_{\beta (0)}\chi^I_u
	|\tau_1|{\bm \Phi}^{\it (u)}_{\beta' (2)}\chi^I_d \rangle 
	\nonumber \\
	&&=\frac{1}{2\pi} \int d^{3}x
	{\phi}^{|\omega_0|}_{n_r}(\rho){\phi}_{n_z}(z) e^{-i\omega_0 \varphi }
	{\phi}^{|\omega_2'|}_{n_r'} (\rho){\phi}_{n_z'} (z) e^{i\omega_2'\varphi}
	\nonumber \\
	&&=\delta_{n_r n_r'}\delta_{n_z n_z'}
	\delta_{K_3-\frac{m_{\rm w}}{2} K_3'+\frac{m_{\rm w}}{2}} 
	\nonumber \\
	&&\langle {\bm \Phi}^{\it (n)}_{\alpha(1)}\chi^I_u
	|\tau_1|{\bm \Phi}^{\it (u)}_{\alpha'(3)}\chi^I_d \rangle
	=\langle {\bm \Phi}^{\it (n)}_{\beta(1)}\chi^I_u
	|\tau_1|{\bm \Phi}^{\it (u)}_{\beta'(3)}\chi^I_d\rangle
	\nonumber 
	\\
	&&= \frac{1}{2 \pi} \int d^{3} x   {\phi}^{|\omega_1|}_{n_r}(\rho)
	{\phi}_{n_z}(z) e^{-i\omega_1 \varphi }
	{\phi}^{|\omega_3'|}_{n_r'} (\rho) 
	{\phi}_{n_z'} (z) e^{i\omega_3'\varphi}
	\nonumber
	\\
	&&=\delta_{n_r n_r'}\delta_{n_z n_z'}
	\delta_{K_3-\frac{m_{\rm w}}{2} K_3'+\frac{m_{\rm w}}{2}} 
	\nonumber \\
	&&\langle {\bm \Phi}^{\it (n)}_{\alpha(2)}\chi^I_d
	|\tau_1|{\bm \Phi}^{\it (u)}_{\alpha'(0)}\chi^I_u\rangle
	=\langle {\bm \Phi}^{\it (n)}_{\beta(2)}\chi^I_d
	|\tau_1|{\bm \Phi}^{\it (u)}_{\beta'(0)}\chi^I_u\rangle
	\nonumber 
	\\
	&&= \frac{1}{2 \pi} \int d^{3} x   {\phi}^{|\omega_2|}_{n_r}(\rho)
	{\phi}_{n_z}(z) e^{-i\omega_2 \varphi }
	{\phi}^{|\omega_0'|}_{n_r'} (\rho) 
	{\phi}_{n_z'} (z) e^{i\omega_0'\varphi}
	\nonumber
	\\
	&&=\delta_{n_r n_r'}\delta_{n_z n_z'}
	\delta_{K_3+\frac{m_{\rm w}}{2} K_3'-\frac{m_{\rm w}}{2}} 
	\nonumber \\
	&&\langle {\bm \Phi}^{\it (n)}_{\alpha(3)}\chi^I_d
	|\tau_1|{\bm \Phi}^{\it (u)}_{\alpha'(1)}\chi^I_u\rangle
	=\langle {\bm \Phi}^{\it (n)}_{\beta(3)}\chi^I_d
	|\tau_1|{\bm \Phi}^{\it (u)}_{\beta'(1)}\chi^I_u\rangle
	\nonumber 
	\\
	&&= \frac{1}{2 \pi} \int d^{3} x   {\phi}^{|\omega_3|}_{n_r}(\rho)
	{\phi}_{n_z}(z) e^{-i\omega_3 \varphi }
	{\phi}^{|\omega_1'|}_{n_r'} (\rho) 
	{\phi}_{n_z'} (z) e^{i\omega_1'\varphi}
	\nonumber
	\\&&=\delta_{n_r n_r'}\delta_{n_z n_z'}
	\delta_{K_3+\frac{m_{\rm w}}{2} K_3'-\frac{m_{\rm w}}{2}} \,.
\end{eqnarray}
For $\langle n| \tau_3|m \rangle$, 
\begin{eqnarray}
	&&\langle {\bm \Phi}_{\alpha(0)}\chi^I_u
	|\tau_3|
	{\bm \Phi}_{\alpha'(0)}\chi^I_u\rangle
	=\langle {\bm \Phi}_{\alpha(1)}\chi^I_u
	|\tau_3|
	{\bm \Phi}_{\alpha'(1)}\chi^I_u\rangle \nonumber \\
	&&=\langle {\bm \Phi}_{\beta(0)}\chi^I_u
	|\tau_3|
	{\bm \Phi}_{\beta'(0)}\chi^I_u\rangle
	=\langle {\bm \Phi}_{\beta(1)}\chi^I_u
	|\tau_3|
	{\bm \Phi}_{\beta'(1)}\chi^I_u\rangle
	\nonumber \\
	&&= \frac{1}{2 \pi} \int d^{3} x   {\phi}^{|\omega_0|}_{n_r}(\rho)
	{\phi}_{n_z}(z) e^{-i\omega_0 \varphi }
	{\phi}^{|\omega_0'|}_{n_r'} (\rho) 
	{\phi}_{n_z'} (z) e^{i\omega_0'\varphi} \nonumber \\
	&&=\delta_{n_r n_r'}\delta_{n_z n_z'}\delta_{K_3 K_3'} 
	\nonumber \\
	&&\langle {\bm \Phi}_{\alpha(2)}\chi^I_d
	|\tau_3|
	{\bm \Phi}_{\alpha'(2)}\chi^I_d\rangle
	=\langle {\bm \Phi}_{\alpha(3)}\chi^I_d
	|\tau_3|
	{\bm \Phi}_{\alpha'(3)}\chi^I_d\rangle \nonumber \\
	&&=\langle {\bm \Phi}_{\beta(2)}\chi^I_d
	|\tau_3|
	{\bm \Phi}_{\beta'(2)}\chi^I_d\rangle
	=\langle {\bm \Phi}_{\beta(3)}\chi^I_d|\tau_3|
	{\bm \Phi}_{\beta'(3)}\chi^I_d\rangle
	\nonumber \\
	&&= -\frac{1}{2 \pi} \int d^{3} x   {\phi}^{|\omega_0|}_{n_r}(\rho)
	{\phi}_{n_z}(z) e^{-i\omega_0 \varphi }
	{\phi}^{|\omega_0'|}_{n_r'} (\rho) 
	{\phi}_{n_z'} (z) e^{i\omega_0'\varphi} \nonumber \\
	&&=-\delta_{n_r n_r'}\delta_{n_z n_z'}\delta_{K_3 K_3'} \,.
\end{eqnarray}
For $\langle n| J_1|m \rangle$, we shall eveluate 
$\langle n |[H(U^{\gamma_5}_0),{J}_1] | m \rangle$ instead 
with the replacement of (\ref{eq:replacement}),   
\begin{eqnarray*}
	&&{\langle n |[H(U^{\gamma_5}_0),{J}_1] | m \rangle} \\
	&&={\langle n |[\beta M(\cos F(\rho,z)+i\gamma_5
	\bm{\tau\cdot}\hat{\bm n}\sin F(\rho,z)) ,{l}_1] | m \rangle} \\
	&&=-{\langle n |l_1[\beta M(\cos F(\rho,z)+i\gamma_5
	\bm{\tau\cdot}\hat{\bm n}\sin F(\rho,z))]| m \rangle} \\
	\label{eq:replace2}
\end{eqnarray*}
where 
\begin{widetext}
\begin{eqnarray}
	&&l_1=-\frac{1}{2}\biggl[e^{i\varphi}\Bigl(\rho\frac{\partial}
	{\partial z}-z\frac{\partial}{\partial \rho}
	-i\frac{z}{\rho}\frac{\partial}{\partial \varphi}\Bigr) 
	-e^{-i\varphi}\Bigl(\rho\frac{\partial}{\partial z}
	-z\frac{\partial}{\partial \rho}
	+i\frac{z}{\rho}\frac{\partial}{\partial \varphi}\Bigr) \biggr] \nonumber \\
	&&l_2=-\frac{1}{2}i\biggl[e^{i\varphi}\Bigl(\rho\frac{\partial}
	{\partial z}-z\frac{\partial}{\partial \rho}
	-i\frac{z}{\rho}\frac{\partial}{\partial \varphi}\Bigr) 
	+e^{-i\varphi}\Bigl(\rho\frac{\partial}{\partial z}-z\frac{\partial}{\partial \rho}
	+i\frac{z}{\rho}\frac{\partial}{\partial \varphi}\Bigr)\biggr] \nonumber \\
	&&l_3=-i\frac{\partial}{\partial \varphi} 
	\label{eq:amomentum}
\end{eqnarray}
are the components of the angular momentum operator $\bm{l}$ 
in cylindrical coordinates. 
Using eqs.(\ref{eq:amomentum}), one obtains 
\begin{eqnarray}
	&&\langle {\bm \Phi}^{\it (n)}_{\alpha(0)}|
	e^{\pm i\varphi}\biggl[\Bigl(\rho\frac{\partial}{\partial z}
	-z\frac{\partial}{\partial \rho}
	\mp i\frac{z}{\rho}\frac{\partial}{\partial \varphi}\Bigr)\cos F(\rho,z)\biggr]
	|{\bm \Phi}^{\it (u)}_{\alpha'(0)}\rangle \nonumber \\
	&&=-\int \rho d\rho dz
	N^{|\omega_0|}_{n_r}N^{|\omega_0'|}_{n_r'}
	(\sqrt{\alpha_\rho})^{|\omega_0|+|\omega_0'|}e^{-\alpha_\rho \rho^2}
	N_{n_z}N_{n_z'}e^{-\alpha_z z^2} \nonumber \\
	&&\hspace{0.5cm}\times\biggl\{\rho^{|\omega_0|+|\omega_0'|+1}
	L^{|\omega_0|}_{n_r}(\alpha_\rho \rho^2)
	L^{|\omega_0'|}_{n_r'}(\alpha_\rho \rho^2)
	\nonumber \\
	&&\hspace{1cm}\times 
	\sqrt{\alpha_z}\biggl[\Bigl(n_zH_{n_z-1}(\sqrt{\alpha_z}z)
	-\frac{1}{2}H_{n_z+1}(\sqrt{\alpha_z}z)\Bigr)
	H_{n_z'}(\sqrt{\alpha_z}z) 
	\nonumber \\
	&&\hspace{1.5cm}+H_{n_z}(\sqrt{\alpha_z}z)\Bigl(n_z'H_{n_z-1}(\sqrt{\alpha_z}z)
	-\frac{1}{2}H_{n_z+1}(\sqrt{\alpha_z}z)\Bigr)\biggr] 
	\nonumber 
	\\
	&&\hspace{1cm}-\rho^{|\omega_0|+|\omega_0'|-1}
	\biggl[\Bigl((2n_r+|\omega_0|-\alpha_\rho \rho^2)
	L^{|\omega_0|}_{n_r}(\alpha_\rho \rho^2)
	-2(n_r+|\omega_0|)L^{|\omega_0|}_{n_r-1}(\alpha_\rho \rho^2)\Bigr)
	L^{|\omega_0'|}_{n_r'}(\alpha_\rho \rho^2) 
	\nonumber \\
	&&\hspace{1.5cm}+L^{|\omega_0|}_{n_r}(\alpha_\rho \rho^2)
	\Bigl((2n_r'+|\omega_0'|-\alpha_\rho \rho^2)L^{|\omega_0'|}_{n_r'}(\alpha_\rho \rho^2)
	-2(n_r'+|\omega_0'|)L^{|\omega_0'|}_{n_r'-1}(\alpha_\rho \rho^2)\Bigr)
	\pm L^{|\omega_0|}_{n_r}(\alpha_\rho \rho^2) 
	L^{|\omega_0'|}_{n_r'}(\alpha_\rho \rho^2)\biggr]
	\nonumber \\
	&&\hspace{1cm}\times 
	H_{n_z}(\sqrt{\alpha_z}z) zH_{n_z'}(\sqrt{\alpha_z}z)\biggr\} \cos F(\rho,z)
	\delta_{K_3\mp \frac{1}{2} K_3'\pm\frac{1}{2}} 
	\label{eq:mati00} 
\end{eqnarray}
\begin{eqnarray}
	&&\langle {\bm \Phi}^{\it (n)}_{\alpha(0)}|
	ie^{\pm i\varphi}\biggl[\Bigl(\rho\frac{\partial}{\partial z}
	-z\frac{\partial}{\partial \rho}
	\mp i\frac{z}{\rho}\frac{\partial}{\partial \varphi} \Bigr)\sin F(\rho,z) 
	\cos \Theta(\rho,z)\biggr]
	|i{\bm \Phi}^{\it (u)}_{\beta'(0)}\rangle \nonumber \\
	&&=\int \rho d\rho dz N^{|\omega_0|}_{n_r}N^{|\omega_0'|}_{n_r'}
	 (\sqrt{\alpha_\rho})^{|\omega_0|+|\omega_0'|}e^{-\alpha_\rho \rho^2}
	N_{n_z}N_{n_z'}e^{-\alpha_z z^2} \nonumber \\
	&&\hspace{0.5cm}\times\biggl\{ \rho^{|\omega_0|+|\omega_0'|+1}
	L^{|\omega_0|}_{n_r}(\alpha_\rho \rho^2)
	L^{|\omega_0'|}_{n_r'}(\alpha_\rho \rho^2) 
	\nonumber \\
	&&\hspace{1cm}\times \sqrt{\alpha_z}\biggl[\Bigl(n_zH_{n_z-1}(\sqrt{\alpha_z}z)
	-\frac{1}{2}H_{n_z+1}(\sqrt{\alpha_z}z)\Bigr) H_{n_z'}(\sqrt{\alpha_z}z) 
	\nonumber \\
	&&\hspace{1.5cm}+H_{n_z}(\sqrt{\alpha_z}z) \Bigl(n_z'H_{n_z-1}(\sqrt{\alpha_z}z)
	-\frac{1}{2}H_{n_z+1}(\sqrt{\alpha_z}z)\Bigr)
	\biggr]
	\nonumber \\
	&&\hspace{1cm}-\rho^{|\omega_0|+|\omega_0'|-1}
	\biggl[\Bigl((2n_r+|\omega_0|-\alpha_\rho \rho^2)
	L^{|\omega_0|}_{n_r}(\alpha_\rho \rho^2)
	-2(n_r+|\omega_0|)L^{|\omega_0|}_{n_r-1}(\alpha_\rho \rho^2)\Bigr)
	L^{|\omega_0'|}_{n_r'}(\alpha_\rho \rho^2) 
	\nonumber \\
	&&\hspace{1.5cm}+L^{|\omega_0|}_{n_r}(\alpha_\rho \rho^2)
	\Bigl((2n_r'+|\omega_0'|-\alpha_\rho \rho^2)L^{|\omega_0'|}_{n_r'}(\alpha_\rho \rho^2)
	-2(n_r'+|\omega_0'|)L^{|\omega_0'|}_{n_r'-1}(\alpha_\rho \rho^2)\Bigr) 
	\pm L^{|\omega_0|}_{n_r}(\alpha_\rho \rho^2) 
	L^{|\omega_0'|}_{n_r'}(\alpha_\rho \rho^2)\biggr]
	\nonumber \\
	&&\hspace{1cm}\times 
	H_{n_z}(\sqrt{\alpha_z}z) zH_{n_z'}(\sqrt{\alpha_z}z) \biggr\}
	 \sin F(\rho,z) \cos\Theta(\rho,z) \delta_{K_3\mp\frac{1}{2} K_3'\pm\frac{1}{2}} 
	\label{eq:mati002}
\end{eqnarray}
\begin{eqnarray}
	&&\langle {\bm \Phi}^{\it (n)}_{\alpha(0)}|
	ie^{\pm i\varphi}\biggl[\Bigl(\rho\frac{\partial}{\partial z}
	-z\frac{\partial}{\partial \rho}
	\mp i\frac{z}{\rho}\frac{\partial}{\partial \varphi}\Bigr)
	\sin F(\rho,z) \sin \Theta(\rho,z)e^{-im\varphi}\biggr]
	|i{\bm \Phi}^{\it (u)}_{\beta'(2)}\rangle \nonumber \\
	&&=\int \rho d\rho dz 
	N^{|\omega_0|}_{n_r}N^{|\omega_2'|}_{n_r'}
	\sqrt{\alpha_\rho})^{|\omega_0|+|\omega_2'|}e^{-\alpha_\rho \rho^2} 
	N_{n_z}N_{n_z'}e^{-\alpha_z z^2} \nonumber \\
	&&\hspace{0.5cm}\times\biggl
	\{\rho^{|\omega_0|+|\omega_2'|+1}
	L^{|\omega_0|}_{n_r}(\alpha_\rho \rho^2)
	L^{|\omega_2'|}_{n_r'}(\alpha_\rho \rho^2) \nonumber \\
	&&\hspace{1cm}\times \sqrt{\alpha_z}
	\biggl[\Bigl(n_zH_{n_z-1}(\sqrt{\alpha_z}z)-\frac{1}{2}H_{n_z+1}(\sqrt{\alpha_z}z)\Bigr)
	H_{n_z'}(\sqrt{\alpha_z}z) 
	\nonumber \\
	&&\hspace{1.5cm}+H_{n_z}(\sqrt{\alpha_z}z)
	\Bigl(n_z'H_{n_z-1}(\sqrt{\alpha_z}z)-\frac{1}{2}H_{n_z+1}(\sqrt{\alpha_z}z)\Bigr)
	\biggr]
	\nonumber \\
	&&\hspace{1cm}-
	\rho^{|\omega_0|+|\omega_2'|-1}
	\biggl[\Bigl((2n_r+|\omega_0|-\alpha_\rho \rho^2)
	L^{|\omega_0|}_{n_r}(\alpha_\rho \rho^2)
	-2(n_r+|\omega_0|)L^{|\omega_0|}_{n_r-1}(\alpha_\rho \rho^2)\Bigr)
	L^{|\omega_2'|}_{n_r'}(\alpha_\rho \rho^2) \nonumber \\
	&&\hspace{1.5cm}+L^{|\omega_0|}_{n_r}(\alpha_\rho \rho^2)
	\Bigl((2n_r'+|\omega_2'|-m-\alpha_\rho \rho^2)
	L^{|\omega_2'|}_{n_r'}(\alpha_\rho \rho^2)
	-2(n_r'+|\omega_2'|)L^{|\omega_2'|}_{n_r'-1}(\alpha_\rho \rho^2)\Bigr)
	\pm L^{|\omega_0|}_{n_r}(\alpha_\rho \rho^2)
	 L^{|\omega_2'|}_{n_r'}(\alpha_\rho \rho^2)\biggr]
	\nonumber \\
	&&\hspace{1cm}\times H_{n_z}(\sqrt{\alpha_z}z) zH_{n_z'}(\sqrt{\alpha_z}z)\biggr\}
	\sin F(\rho,z) \sin\Theta(\rho,z) \delta_{K_3\mp\frac{1}{2} K_3'\pm\frac{1}{2}}.
\end{eqnarray}
Other elements can be obtained in the same manner. 
In Appendix \ref{sec:levelb}, we shall present these calculations 
in more detail. 
\end{widetext}

\subsection{ Numerical Convergence}
In this subsection, we show the convergence of the 
soliton energy $E_{static}$ with respect to ‡T) $K_3$ with the basis 
number fixed, and ‡U) the number of the basis (discretized momentum number) 
with $K_3$ fixed.

\begin{figure}
	\includegraphics[height=7cm, width=9cm]{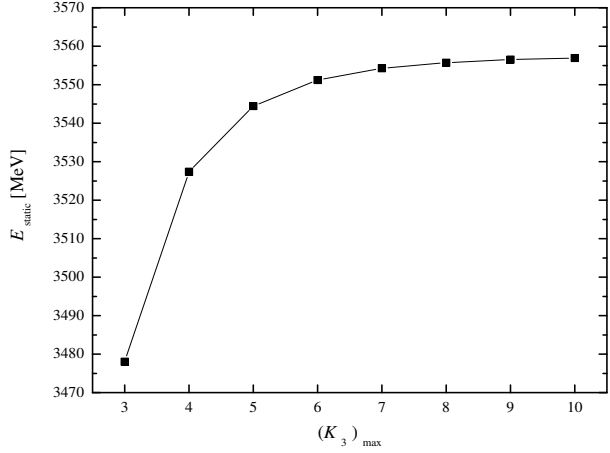}
	\caption{\label{fig:Cpq_k} Convergence of $(K_3)_{max}$-$E_{static}$ 
	for $B=3$ calculation with $N_{rmax}=N_{zmax}=10$.}
\end{figure}

Fig.\ref{fig:Cpq_k} shows the case of ‡T) with $B=3$. 
As can be seen, the energy is almost convergent at the $(K_3)_{max}=10$. 
\begin{figure}
	\includegraphics[height=7cm, width=9cm]{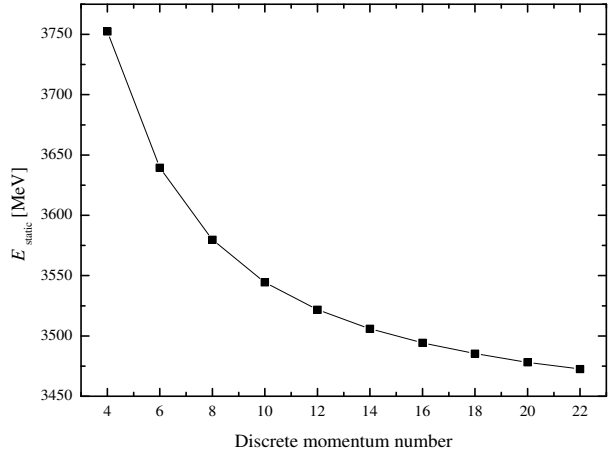}
	\caption{\label{fig:Cpq_e} Convergence for the basis number - $E_{static}$.
	$(K_3)_{max}$ is fixed by $6$.}
\end{figure}
In the case of ‡U), the energy is not perfectly convergent 
up to $N_{rmax}=N_{zmax}=22$~(see Fig.\ref{fig:Cpq_e}). 
Therefore, all our results have $1-2~\%$ uncertainty.    

\begin{table}
	\caption{\label{tab:b2}$B=2$, mass spectrum up to $i,j\le 3, k_3\le 1$.}
	\begin{ruledtabular}
	\begin{tabular}{cccc}
	~~Classification & ($i,j,k_3$) & Parity & Mass[$MeV$] \\
	\hline
	$NN(^{3}{S}_{1})$              &($ 0,1,0$)& $+$ &2264\\
	$NN(^{1}{S}_{0})$              &($ 1,0,0$)& $+$ &2290\\
	$N\Delta(^{3}{P}_{2})$         &($ 1,2,1$)& $-$ &2399\\
	$N \Delta (^{5}{S}_{2})$       &($ 1,2,0$)& $+$ &2477\\
	$N \Delta (^{3}{S}_{1})$       &($ 2,1,0$)& $+$ &2528\\
	$ \Delta \Delta (^{7}{S}_{3})$ &($ 0,3,0$)& $+$ &2576\\
	$ \Delta \Delta (^{1}{S}_{0})$ &($ 3,0,0$)& $+$ &2730\\
	$\Delta\Delta(^{5}{P}_{3})$    &($ 2,3,1$)& $-$ &2762\\
	\end{tabular}
	\end{ruledtabular}
\end{table}

\begin{table}
	\caption{\label{tab:b4}$B=4$, mass spectrum up to $i\le 3,j\le 5,k_3\le 1$.}
	\begin{ruledtabular}
	\begin{tabular}{cccc}
	~~Classification & ($i,j,k_3$) & Parity & Mass[$MeV$] \\
	\hline
	$^{4}N(^{1}{S}_{0})$              &($ 0,0,0$)& $+$ &4753\\
	$^{4}N(^{5}{S}_{2})$              &($ 0,2,0$)& $+$ &4807\\
	$^{2}N~^{2}\Delta(^{7}{P}_{4})$   &($ 0,4,1$)& $-$ &4808\\
	$^{4}N(^{3}{S}_{1})$              &($ 1,1,0$)& $+$ &4813\\
	$^{4}N(^{1}{S}_{0})$              &($ 2,0,0$)& $+$ &4879\\
	$^{3}N~\Delta(^{7}{S}_{3})$       &($ 1,3,0$)& $+$ &4904\\
	$^{4}N(^{5}{S}_{2})$              &($ 2,2,0$)& $+$ &4934{\tiny.2}\\
	$^{2}N~^{2}\Delta(^{7}{S}_{4})$   &($ 0,4,0$)& $+$ &4934{\tiny.3}\\
	$^{2}N~^{2}\Delta(^{9}{P}_{4})$   &($ 2,4,1$)& $-$ &4935\\
	$N~^{3}\Delta(^{9}{P}_{5})$       &($ 1,5,1$)& $-$ &4941\\
	$^{3}N~\Delta(^{3}{S}_{1})$       &($ 3,1,0$)& $+$ &5025\\
	$N~^{3}\Delta(^{9}{S}_{5})$       &($ 1,5,0$)& $+$ &5046\\
	$^{2}N~^{2}\Delta(^{9}{S}_{4})$   &($ 2,4,0$)& $+$ &5061\\
	$^{3}N~\Delta(^{7}{S}_{3})$       &($ 3,3,0$)& $+$ &5115\\
	$N~^{3}\Delta(^{9}{P}_{5})$       &($ 3,5,1$)& $-$ &5152\\
	$N~^{3}\Delta(^{9}{S}_{5})$       &($ 3,5,0$)& $+$ &5278\\
	\end{tabular}
	\end{ruledtabular}
\end{table}

\section{\label{sec:level5}CONCLUSIONS}
In this paper, we studied axially symmetric soliton solutions 
with $B=2-5$ in the chiral quark soliton model. 
The one-quark spectral flow analysis indicates 
that the number of diving states from positive continuum to negative 
coincide with the baryon number. 
As is shown in Table \ref{tab:claspectra}, 
the valence quark spectra contain double degeneracy, 
realisng lower energy than non-degenerate states. 
Therefore, our solitons are stable although they are not 
necessarily minimal energy one, which confirms that   
they are good saddle point solutions. 

Upon quantization, we computed zero-mode rotational corrections to 
the classical energy. The study of the Finkelstein-Rubinstein 
constraints indicates that the axially symmetric solution with 
even $B$ has the same quantum number as the physically observed 
nuclei. 
These results are shown in Table \ref{tab:b2} and \ref{tab:b4}. 
Some of the states may be observed in experiments. 
For odd $B$, the constraint of $C_2$ in Eq.(\ref{j1}) seems 
to assure the validity of the ansatz. Indeed, it provides 
the ground state as $i=j=1/2$ for $B=3$ and as $i=1/2, j=3/2$ for $B=5$, 
which exactly coincide with physical observations. This seems to make 
sense since in the minimal energy configurations with discrete symmetries, 
the solutions tend to have $i=j=1/2$ due to their shell-like 
structure. However, unfortunately the constraint in Eq.(\ref{j3}) 
forbit such states. 
Consequently, the axially symmetric solitons with odd $B$ emerge 
only as excited states. 
The resultant lowest state is $E=3657$ MeV 
with $i=1/2,j=5/2$ for $B=3$, and  is $E=6591$ MeV with $i=1/2,j=7/2$ 
for $B=5$.
 
Recently, we also studied classical multi-baryonic systems 
with discrete symmetries in the CQSM and found larger 
degeneracy of the quark orbits than of the axially symmetric 
\cite{sawado02,sawado03}. 
For example, triply degenerate bound spectrum is 
obtained in the $B=3$ tetrahedral soliton background.
The interesting point is that the corresponding 
energy of the soliton is $E\sim 210$ MeV which is higher
than the axially symmetric, $E= 173$ MeV 
(see Table \ref{tab:claspectra}). Likewise, 
for the $B=4$ minimal energy soliton with cubic 
symmetry, the valence quark spectrum shows four-fold degeneracy
with $E\sim 170$ MeV while for the axially symmetric, 
$E=106, 232$ MeV.  
Thus, although the degeneracy of the spectum indicates 
the stability of the solutions, other factors
should be also taken into account in regard to minimization 
of their classical energies.  
More detailed discussions on this subject will be made elsewhere. 

\begin{acknowledgments}
We thank S.Oryu for useful discussions. 
We are also grateful to N.S.Manton to inform the paper of S.Krusch
(ref.\cite{krusch}). 
One of us (Sawado) also thanks M.Kawabata and K.Saito for their 
help of numerical computations. 
\end{acknowledgments}

\appendix

\section{\label{sec:levela}Evaluation of the matrix elements}

In this appendix, we shall present detailed calculations of 
the matrix elements, $\int d^{3}x {\phi}_{\mu}^{\dagger} H {\phi}_{\nu} $.

For the kinetic term of the elements, one gets 
\begin{eqnarray}
	&& \langle {\bm \Phi}_{\alpha(0)}|{\bm{\sigma \cdot p}}
	|i{\bm \Phi}_{\beta'(0)}\rangle \nonumber \\
	&&= \frac{1}{2 \pi} \int d^{3} x
	\phi^{|\omega_0|}_{n_r}(\rho) \phi_{n_z}(z) e^{-i\omega_0 \varphi }
	\nonumber \\
	&&\hspace{0.5cm}\times\Bigl(\frac{\partial}{\partial z}\Bigr)
	 {\phi}^{|\omega_0'|}_{n_r'}(\rho){\phi}_{n_z'} (z)
	 e^{i\omega_0 \varphi }  \nonumber
	\nonumber \\
	&&=\int \rho d \rho dz {\phi}^{|\omega_0|}_{n_{r}} {\phi}^{|\omega_0|}_{n_{r'}}
	 N_{n_z} e^{-\alpha_z z^2} H_{n_z} (\sqrt{\alpha_z} z) 
	\nonumber \\ 
	&&\hspace{0.5cm}\times N_{n_z'}
	\sqrt{\alpha_z}\Bigl(n_z' H_{n_z'- 1}(\sqrt{\alpha_z} z)
	- \frac{1}{2} H_{n_z'+ 1}(\sqrt{\alpha_z} z)\Bigr)
	\nonumber \\
	&&=\delta_{n_r n_r'}\delta_{\omega_{0} \omega_{0'}}
	N_{n_z} N_{n_z'}\frac{1}{N^{2}_{n_z}} 
	\nonumber \\
	&&\hspace{2cm}\times\sqrt{\alpha_z}\Bigl(n_z' \delta_{n_z n_z'-1}
	- \frac{1}{2} \delta_{n_z n_{z'+1}}\Bigr) \,.
	\nonumber \\
	\label{eq:a1}
\end{eqnarray}
For $(n_z,n_z')=(1,2),(3,4), \cdots$ or $(n_z,n_z')=(0,1),(2,3), \cdots$, 
the righthand side of Eq.(\ref{eq:a1}) becomes
\begin{eqnarray}
	= \delta_{n_r n_r'}\delta_{\omega_0 \omega_0'}
	\frac{N_{n_z'}}{N_{n_z}}\sqrt{\alpha_z}
	n_z'\delta_{n_z n_z'-1} \,.
	\label{eq:a5}
\end{eqnarray}
For $(n_z,n_z')=(1,0),(3,2), \cdots$ or $(n_z,n_z')=(2,1),(4,3), \cdots$,   
\begin{eqnarray}
	=-\frac{1}{2}\delta_{n_r n_r'}\delta_{\omega_0 \omega_0'}
 	\frac{N_{n_z'}}
	{N_{n_z}} \sqrt{\alpha_z} n_z' \delta_{n_r n_r'+1}\,.
	\label{eq:a6}
\end{eqnarray}
Also, 
\begin{eqnarray}
	&& \langle {\bm \Phi}_{\alpha(0)}|{\bm{\sigma \cdot p}}|i{\bm \Phi}_{\beta'(1)}\rangle
	\nonumber
	 \\
	&&= \frac{1}{2 \pi} \int d^3 x   \phi^{|\omega_0|}_{n_r}(\rho)
	 {\phi}_{n_z}(z) e^{-i\omega_0 \varphi }e^{-i \varphi}
	\nonumber
	\\
	&&\hspace{0.5cm}\times \Bigl(\frac{\partial}{\partial \rho}
	 - \frac{i}{\rho} \frac{\partial}
	{\partial \varphi}\Bigr) {\phi}^{|\omega_1'|}_{n_r'} (\rho) 
	{\phi}_{n_z'} (z) e^{i\omega_1 \varphi } 
	\nonumber 
	\\
	&&=\delta_{\omega_0 \omega_1'} \delta_{n_z n_z'} \int \rho d\rho
	 {\phi}^{|\omega_0|}_{n_r} 
	\Bigl(\frac{\partial}{\partial \rho} 
	+ \frac{\omega_1'}{\rho}\Bigr)
	{\phi}^{|\omega_1'|}_{n_r'}
	\nonumber 
	\\
	&&=\delta_{\omega_0 \omega_1'} \delta_{n_z n_z'}
	\int \rho d \rho  (\sqrt{\alpha_\rho})^{|\omega_0|+|\omega_1'|}
	{\rho}^{|\omega_0|+|\omega_1'|-1}e^{-\alpha_\rho \rho^2} 
	\nonumber 
	\\
	&&\hspace{0.5cm}\times N^{|\omega_0|}_{n_z} N^{|\omega_1'|}_{n_z'}L^{|\omega_0|}_{n_r}
	(\alpha_\rho \rho^2)
	\Bigl((2 n_r' + |\omega_1'| + \omega_1' - \alpha \rho^2)
	\nonumber 
	\\
	&&\hspace{0.5cm}\times L^{|\omega_1'|}_{n_r'}(-\alpha_\rho \rho^2) 
	-2(n_r'+|\omega_1'|) L^{|\omega_1'|}_{n_r'-1}(\alpha \rho^2)\Bigr)\,.
	\label{ap:a4}
\end{eqnarray}
For $\omega_0 \ge 0$ $(\omega_1'=\omega_0+1>0)$, the righthand side of 
Eq.(\ref{ap:a4}) becomes
\begin{eqnarray}
	&&={\delta}_{n_z n_z'}\int \rho d \rho  (\sqrt{\alpha_\rho})^{2\omega_0+1}
	\rho^{2\omega_0}e^{-\alpha_\rho \rho^2}N^{\omega_0}_{n_r}
	N^{\omega_0+1}_{n_r'} 
	\nonumber \\
	&&\hspace{0.5cm}\times L^{\omega_0}_{n_r}(\alpha_\rho \rho^2)
	(2 n_r' + 2\omega_0 + 2) 
	L^{\omega_0}_{n_r'}(\alpha_\rho \rho^2) \nonumber
	\\
	&&\hspace{0.5cm}-{\delta}_{n_z n_z'}\int \rho d \rho  (\sqrt{\alpha_\rho})^{2\omega_0+3}
	{\rho}^{2\omega_0+2}e^{-\alpha_\rho \rho^2}
	N^{\omega_0}_{n_z} N^{\omega_0+1}_{n_z'}
	\nonumber \\
	&&\hspace{0.5cm}\times \Bigl( L^{\omega_0+1}_{n_r}
	(\alpha_\rho {\rho}^{2})-L^{\omega_0+1}_{n_r-1}(\alpha_\rho \rho^2)\Bigr) 
	L^{\omega_0+1}_{n_r'}(\alpha_\rho \rho^2) \nonumber
	\\
	&&=\delta_{n_z n_z'}\sqrt{\alpha_\rho} N^{\omega_0}_{n_r}
	N^{\omega_0+1}_{n_r'} (2n_r'+2\omega_0+2) 
	\frac{1}{N^{\omega_0 2}_{n_r}}{\delta}_{n_r n_r'}
	\nonumber \\
	&&\hspace{0.5cm}-\delta_{n_z n_z'}{\sqrt{\alpha_\rho}}N^{\omega_0}_{n_r}
	N^{\omega_0+1}_{n_r'}
	\nonumber \\
	&&\hspace{1cm}
	 \times\Bigl(\frac{1}{N^{\omega_0+1 2}_{n_r}}{\delta}_{n_r n_r'}-
	\frac{1}{N^{\omega_0+1 2}_{n_r}}{\delta}_{n_r-1 n_r'}\Bigr) \nonumber
	\\
	&&=\delta_{n_z n_z'}\sqrt{\alpha_\rho}
	(2 \sqrt{n_r+\omega_0+1}\delta_{n_r n_r'}
	\nonumber \\
	&&\hspace{0.5cm}-\sqrt{n_r+\omega_0+1}\delta_{n_r n_r'}
	+\sqrt{n_r}\delta_{n_r-1 n_r'}) \nonumber
	\\
	&&=\delta_{n_z n_z'}\sqrt{\alpha_\rho}
	(\sqrt{n_r+\omega_0+1}\delta_{n_r n_r'}
	+\sqrt{n_r}\delta_{n_r-1 n_r'})
\end{eqnarray} 
where
\begin{eqnarray}
	&&N^{\omega_0}_{n_r}N^{\omega_0+1}_{n_r'}
	\frac{\delta_{n_r n_r'}}{N^{\omega_0 2}_{n_r}} 
	=\frac{\delta_{n_r n_r'}}{\sqrt{n_r+\omega_0+1}},\\
	&&N^{\omega_0}_{n_r}N^{\omega_0+1}_{n_r'}
	\frac{\delta_{n_r n_r'}}{N^{\omega_0+1 2}_{n_r}} 
	=\sqrt{n_r+\omega_0+1}\delta_{n_r n_r'},\\
	&&N^{\omega_0}_{n_r}N^{\omega_0+1}_{n_r'}
	\frac{\delta_{n_r-1 n_r'}}{N^{\omega_0+1 2}_{n_r'}} 
	=\sqrt{n_r}\delta_{n_r-1 n_r'}.
\end{eqnarray} 
have been used. 

Likewise, for $\omega_0 < 0$ $(\omega_1'=\omega_0+1 \le 0)$, 
\begin{eqnarray}
	&&={\delta}_{n_z n_z'}\int \rho d \rho  (\sqrt{\alpha_\rho})^{-2\omega_0-1}
	{\rho}^{-2\omega_2-2}e^{-\alpha_\rho \rho^2}
	\nonumber \\
	&&\hspace{0.5cm}\times N^{-\omega_0}_{n_r}{N}^{-\omega_0-1}_{n_r'}
	 {L}^{-\omega_0}_{n_r}
	(\alpha_\rho {\rho}^{2}) \rho \frac{d}{d \rho} 
	L^{-\omega_0-1}_{n_{r'}}(\alpha_\rho {\rho}^{2}) 
	\nonumber
	\\
	&&\hspace{0.5cm}-{\delta}_{n_z n_z'}\int \rho d \rho  (\sqrt{\alpha_\rho})^{-2\omega_0+1}
	\rho^{-2\omega_0}e^{-\alpha_\rho \rho^2}
	\nonumber \\
	&&\hspace{1cm}\times {N}^{-\omega_0}_{n_r} N^{-\omega_0-1}_{n_r'} 
	L^{-\omega_0}_{n_r}(\alpha_\rho {\rho}^{2}) 
	L^{-\omega_0-1}_{n_r'}(\alpha_\rho {\rho}^{2}) 
	\nonumber
	\\
	&&=\delta_{n_z n_z'}\int \rho d \rho 
	 (\sqrt{\alpha_\rho})^{-2\omega_0-1}
	\rho^{-2\omega_2-2}e^{-\alpha_\rho \rho^2}
	\nonumber 
	\\
	&&\hspace{0.5cm}\times N^{-\omega_0}_{n_r}{N}^{-\omega_0-1}_{n_r'}
	{L}^{-\omega_0}_{n_r}
	(\alpha_\rho {\rho}^{2}) (-2\alpha\rho^2)
	L^{-\omega_0-1}_{n_{r'}}(\alpha_\rho {\rho}^{2}) 
	\nonumber
	\\
	&&\hspace{0.5cm}-{\delta}_{n_z n_z'}\int \rho d \rho  (\sqrt{\alpha_\rho})^{-2\omega_0+1}
	\rho^{-2\omega_0}e^{-\alpha_\rho \rho^2}
	\nonumber \\
	&&\hspace{1cm}\times N^{-\omega_0}_{n_r}N^{-\omega_0-1}_{n_r'} 
	L^{-\omega_0}_{n_{r}}(\alpha_\rho {\rho}^{2})
	\nonumber \\
	&&\hspace{1.5cm}\times \Bigl(L^{-\omega_0}_{n_r'} (\alpha_\rho {\rho}^{2})
	-L^{-\omega_0}_{n_r'-1}(\alpha_\rho {\rho}^{2})\Bigr) \nonumber
	\\
	&&=-2{\delta}_{n_z n_z'}\sqrt{\alpha_\rho}N^{-\omega_0}_{n_r}
	N^{-\omega_0-1}_{n_r'}\frac{1}{N^{-\omega_0 ~2}_{n_r}}
	\delta_{n_r n_r'-1}
	\nonumber \\
	&&\hspace{0.5cm}-{\delta}_{n_z n_z'}{\sqrt{\alpha_\rho}}{N}^{-\omega_0}_{n_r}
	N^{-\omega_0-1}_{n_r'}
	\nonumber \\
	&&\hspace{1cm}\times \Bigl(\frac{1}{N^{-\omega_0 2}_{n_{r}}}{\delta}_{n_r n_r'}-
	\frac{1}{N^{-\omega_0 2}_{n_{r}}}{\delta}_{n_r n_r'-1}\Bigr) \nonumber
	\\
	&&=-\delta_{n_z n_z'}{\sqrt{\alpha_\rho}}(2 \sqrt{n_r+1}
	\delta_{n_r n_r'-1} 
	\nonumber \\
	&&\hspace{0.5cm}+\sqrt{n_r-\omega_0}{\delta}_{n_r n_r'} -\sqrt{n_r+1}
	\delta_{n_r n_r'-1}) \nonumber
	\\
	&&=-\delta_{n_z n_z'} \sqrt{\alpha_\rho}(\sqrt{n_r-\omega_0} 
	\delta_{n_r n_r'} +\sqrt{n_r+1}{\delta}_{n_r n_r'-1})
	\nonumber \\
\end{eqnarray} 
where 
\begin{eqnarray}
	&&N^{-\omega_0}_{n_r} N^{-\omega_0-1}_{n_r'}\frac{\delta_{n_r n_r'-1}}
	{N^{-\omega_0 2}_{n_r}} = \sqrt{n_r+1}\delta_{n_r n_r'-1}, 
	\\
	&&N^{-\omega_0}_{n_r} N^{-\omega_0-1}_{n_r'}\frac{\delta_{n_r n_r'}}
	{N^{-\omega_0 2}_{n_r}} =\sqrt{n_r-\omega_0}\delta_{n_r n_r'},
	\\
	&&N^{-\omega_0}_{n_r} N^{-\omega_0-1}_{n_r'}\frac{\delta_{n_r n_r'-1}}
	{N^{-\omega_0 2}_{n_r'}} =\sqrt{n_r+1}\delta_{n_r n_r'-1} 
\end{eqnarray} 
have been used. 
In the same manner, elements,
\begin{eqnarray}
	&&\langle {\bm \Phi}_{\alpha(1)}|{\bm{\sigma \cdot p}}
	|i{\bm \Phi}_{\beta'(0)}\rangle
	\langle {\bm \Phi}_{\alpha(1)}|{\bm{\sigma \cdot p}}
	|i{\bm \Phi}_{\beta'(1)}\rangle \nonumber \\
	&&\langle {\bm \Phi}_{\alpha(2)}|{\bm{\sigma \cdot p}}
	|i{\bm \Phi}_{\beta'(2)}\rangle
	\langle {\bm \Phi}_{\alpha(2)}|{\bm{\sigma \cdot p}}
	|i{\bm \Phi}_{\beta'(3)}\rangle \nonumber \\
	&&\cdots 
\end{eqnarray}
can be analytically calculated. 

Since $\bm{\sigma \cdot p}$ does not contain isospin operators, 
following elements vanish 
\begin{eqnarray}
	&&\langle {\bm \Phi}_{\alpha(0)}|{\bm{\sigma \cdot p}}
	|i{\bm \Phi}_{\beta'(2)}\rangle
	\langle {\bm \Phi}_{\alpha(0)}|{\bm{\sigma \cdot p}}
	|i{\bm \Phi}_{\beta'(3)}\rangle \nonumber \\
	&&\langle {\bm \Phi}_{\alpha(1)}|{\bm{\sigma \cdot p}}
	|i{\bm \Phi}_{\beta'(2)}\rangle
	\langle {\bm \Phi}_{\alpha(1)}|{\bm{\sigma \cdot p}}
	|i{\bm \Phi}_{\beta'(3)}\rangle \nonumber \\
	&&\cdots \,.
\end{eqnarray}
For the potential term in the Hamiltonian given by 
\begin{eqnarray}
	M\left( 
	\begin{array}{cc}
	\cos F(\rho,z) &  i \bm{\tau \cdot}  \hat{\bm{n}} \sin F(\rho,z) \\
	-i  \bm{\tau \cdot} \hat{\bm{n}} \sin F(\rho,z) & - \cos F(\rho,z)
	\nonumber
	\end{array}\right)\,,
\end{eqnarray}
only following matrix elements survive 
\begin{eqnarray}
	&&\langle {\bm \Phi}_{\alpha(0)}| M \cos F(\rho,z)
	| {\bm \Phi}_{\alpha'(0)}\rangle 
	\nonumber
	\\
	&&= \int \rho d \rho dz M \cos F(\rho,z)
	\nonumber
	\\
	&&\hspace{0.5cm}\times {\phi}^{|\omega_0|}_{n_r} (\rho) {\phi}_{n_{z}}(z)
	{\phi}^{|\omega_0|}_{{n'}_{r}} (\rho) {\phi}_{{n'}_{z}} (z) \\
	&&\langle {\bm \Phi}_{\alpha(0)}| 
	iM\bm{\tau} \bm{\cdot} \hat{\bm{n}} \sin F(\rho,z)|
	{\bm \Phi}_{\beta'(0)}\rangle 
	\nonumber \\
	&&= -\int \rho d \rho dz M \cos {\Theta}(\rho,z) \sin F(\rho,z) 
	\nonumber
	\\
	&&\hspace{0.5cm}\times {\phi}^{|\omega_0|}_{n_r} (\rho) {\phi}_{n_z}(z)
	{\phi}^{|\omega_0|}_{n_r'} (\rho) {\phi}_{n_z'} (z) \\
	&&\langle {\bm \Phi}_{\alpha(0)}| 
	iM\bm{\tau} \bm{\cdot} \hat{\bm{n}} \sin F(\rho,z)|
	{\bm \Phi}_{\beta'(2)}\rangle 
	\nonumber \\
	&&= -\int \rho d \rho dz M \sin {\Theta}(\rho,z) \sin F(\rho,z) 
	\nonumber
	\\
	&&\hspace{0.5cm}\times {\phi}^{|\omega_0|}_{n_r} (\rho) {\phi}_{n_z}(z)
	{\phi}^{|\omega_2|}_{n_r'} (\rho) {\phi}_{n_z'} (z) \,. 
\end{eqnarray} 
Other elements of the potential term can be calculated in the same manner. 

\section{\label{sec:levelb}Evaluation of the moments of inertia}
\begin{widetext}
In this appendix, we shall give the detailed derivation to obtain  
matrix elements in Eq.(\ref{eq:mati00}).  

For the first term in Eq.(\ref{eq:mati00}), one obtains 
\begin{eqnarray}
	&&\langle {\bm \Phi}^{\it (n)}_{\alpha(0)}|
	e^{\pm i\varphi}\biggl(\rho\frac{\partial}{\partial z}\cos F(\rho,z)\biggr)
	|{\bm \Phi}^{\it (u)}_{\alpha'(0)}\rangle \nonumber \\
	&&=\frac{1}{2\pi}\int d^3x \phi^{|\omega_0|}_{n_r}(\rho)\phi_{n_z}(z) 
	e^{-i\omega_0 \varphi}e^{\pm i\varphi}\biggl(\rho 
	\frac{\partial}{\partial z}\cos F(\rho,z) \biggr)
	\phi^{|\omega_0'|}_{n_r'}(\rho)\phi_{n_z'}(z)e^{i\omega_0'\varphi}
	\nonumber \\
	&&=\int \rho d\rho dz \phi^{|\omega_0|}_{n_r}(\rho)\phi_{n_z}(z)
	\biggl(\rho \frac{\partial}{\partial z}\cos F(\rho,z) \biggr)
	\phi^{|\omega_0'|}_{n_r'}(\rho)\phi_{n_z'}(z)\delta_{K_3\mp\frac{1}{2} K_3'\pm\frac{1}{2}} 
	\label{eq:mati00_0} \,.
\end{eqnarray}
Performing partial integration with respect to $z$ in Eq.(\ref{eq:mati00_0}), 
\begin{eqnarray}
	&&=-\int \rho d\rho dz \phi^{|\omega_0|}_{n_r}(\rho)\rho \phi^{|\omega_0'|}_{n_r'}(\rho) 
	\frac{\partial}{\partial z}\Bigl(\phi_{n_z}(z)\phi_{n_z'}(z)\Bigr)
	\cos F(\rho,z)\delta_{K_3\mp \frac{1}{2} K_3'\pm\frac{1}{2}} 
	\nonumber \\
	&&=-\int \rho d\rho dz N^{|\omega_0|}_{n_r}N^{|\omega_0'|}_{n_r'}
	(\sqrt{\alpha_\rho})^{|\omega_0|+|\omega_0'|}\rho^{|\omega_0|+|\omega_0'|+1}
	e^{-\alpha_\rho \rho^2} L^{|\omega_0|}_{n_r}(\alpha_\rho \rho^2)
	L^{|\omega_0'|}_{n_r'}(\alpha_\rho \rho^2)
	\nonumber \\
	&&\hspace{0.5cm}\times N_{n_z}N_{n_z'}\sqrt{\alpha_z}e^{-\alpha_z z^2}
	\biggl[
	\Bigl(n_zH_{n_z-1}(\sqrt{\alpha_z}z)-\frac{1}{2}H_{n_z+1}(\sqrt{\alpha_z}z)\Bigr)
	H_{n_z'}(\sqrt{\alpha_z}z) 
	\nonumber \\
	&&\hspace{1cm}+H_{n_z}(\sqrt{\alpha_z}z)
	\Bigl(n_z'H_{n_z-1}(\sqrt{\alpha_z}z)-\frac{1}{2}H_{n_z+1}(\sqrt{\alpha_z}z)\Bigr)
	\biggr]\cos F(\rho,z) \delta_{K_3\mp\frac{1}{2} K_3'\pm\frac{1}{2}} \,.
\end{eqnarray}
Similarly, performing partial integration with respect to $\rho$, 
one obtains 
\begin{eqnarray}
	&&\langle {\bm \Phi}^{\it (n)}_{\alpha(0)}|e^{\pm i\varphi}
	\Bigl(z\frac{\partial}{\partial \rho}\cos F(\rho,z)\Bigr)
	|{\bm \Phi}^{\it (u)}_{\alpha'(0)}\rangle \nonumber \\
	&&=\frac{1}{2\pi}\int d^3x \phi^{|\omega_0|}_{n_r}(\rho)\phi_{n_z}(z)
	e^{-i\omega_0 \phi}e^{\pm i\varphi}
	\Bigl(z\frac{\partial}{\partial \rho}\cos F(\rho,z)\Bigr)
	\phi^{|\omega_0'|}_{n_\rho'}(\rho)\phi_{n_z'}(z)e^{i\omega_0'\varphi}
	\nonumber \\
	&&=\int \rho d\rho dz \phi^{|\omega_0|}_{n_r}(\rho)\phi_{n_z}(z)
	\Bigl(z\frac{\partial}{\partial \rho}\cos F(\rho,z) \Bigr)
	\phi^{|\omega_0'|}_{n_\rho'}(\rho)\phi_{n_z'}(z) \delta_{K_3\mp\frac{1}{2} k_3'\pm\frac{1}{2}}
	\nonumber \\
	&&=-\int  d\rho dz \frac{\partial}{\partial \rho} \Bigl(\rho 
	\phi^{|\omega_0|}_{n_r}(\rho)\phi^{|\omega_0'|}_{n_r}(\rho) \Bigr)
	\phi_{n_z}(z)z \phi_{n_z'}(z)\cos F(\rho,z) \delta_{K_3\mp\frac{1}{2} K_3'\pm\frac{1}{2}}
	\nonumber \\
	&&=-\int \rho d\rho dz N^{|\omega_0|}_{n_r}N^{|\omega_0'|}_{n_r'}
	(\sqrt{\alpha_\rho})^{|\omega_0|+|\omega_0'|}\rho^{|\omega_0|+|\omega_0'|-1}
	e^{-\alpha_\rho \rho^2} \nonumber \\
	&&\hspace{0.5cm}\times\biggl[\Bigl((2n_r+|\omega_0|-\alpha_\rho \rho^2)
	L^{|\omega_0|}_{n_r}(\alpha_\rho \rho^2)
	-2(n_r+|\omega_0|)L^{|\omega_0|}_{n_r-1}(\alpha_\rho \rho^2)\Bigr)
	L^{|\omega_0'|}_{n_r'}(\alpha_\rho \rho^2) \nonumber \\
	&&\hspace{1cm}+L^{|\omega_0|}_{n_r}(\alpha_\rho \rho^2)
	\Bigl((2n_r'+|\omega_0'|-\alpha_\rho \rho^2)L^{|\omega_0'|}_{n_r'}(\alpha_\rho \rho^2)
	-2(n_r'+|\omega_0'|)L^{|\omega_0'|}_{n_r'-1}(\alpha_\rho \rho^2)\Bigr)\biggr]
	\nonumber \\
	&&\hspace{0.5cm}\times N_{n_z}N_{n_z'}e^{-\alpha_z z^2}
	H_{n_z}(\sqrt{\alpha_z}z) zH_{n_z'}(\sqrt{\alpha_z}z) \cos F(\rho,z) 
	\delta_{K_3\mp\frac{1}{2} K_3'\pm\frac{1}{2}}\,.
\end{eqnarray}
Finally,  we reach the final answer
\begin{eqnarray}
	&&\langle {\bm \Phi}^{\it (n)}_{\alpha(0)}|
	e^{\pm i\varphi}\biggl[\Bigl(\rho\frac{\partial}{\partial z}-z\frac{\partial}{\partial \rho}
	\mp i\frac{z}{\rho}\frac{\partial}{\partial \varphi}\Bigr)\cos F(\rho,z)\biggr]
	|{\bm \Phi}^{\it (u)}_{\alpha'(0)}\rangle \nonumber \\
	&&=-\int \rho d\rho dz
	N^{|\omega_0|}_{n_r}N^{|\omega_0'|}_{n_r'}
	(\sqrt{\alpha_\rho})^{|\omega_0|+|\omega_0'|}e^{-\alpha_\rho \rho^2}
	N_{n_z}N_{n_z'}e^{-\alpha_z z^2} \nonumber \\
	&&\hspace{0.5cm}\times\biggl\{\rho^{|\omega_0|+|\omega_0'|+1}
	L^{|\omega_0|}_{n_r}(\alpha_\rho \rho^2)
	L^{|\omega_0'|}_{n_r'}(\alpha_\rho \rho^2)
	\nonumber \\
	&&\hspace{1cm}\times 
	\sqrt{\alpha_z}\biggl[\Bigl(n_zH_{n_z-1}(\sqrt{\alpha_z}z)
	-\frac{1}{2}H_{n_z+1}(\sqrt{\alpha_z}z)\Bigr)
	H_{n_z'}(\sqrt{\alpha_z}z) 
	\nonumber \\
	&&\hspace{1.5cm}+H_{n_z}(\sqrt{\alpha_z}z)\Bigl(n_z'H_{n_z-1}
	(\sqrt{\alpha_z}z)-\frac{1}{2}H_{n_z+1}(\sqrt{\alpha_z}z)\Bigr)\biggr] 
	\nonumber \\
	&&\hspace{1cm}-\rho^{|\omega_0|+|\omega_0'|-1}
	\biggl[\Bigl((2n_r+|\omega_0|-\alpha_\rho \rho^2)
	L^{|\omega_0|}_{n_r}(\alpha_\rho \rho^2)
	-2(n_r+|\omega_0|)L^{|\omega_0|}_{n_r-1}(\alpha_\rho \rho^2)\Bigr)
	L^{|\omega_0'|}_{n_r'}(\alpha_\rho \rho^2) 
	\nonumber \\
	&&\hspace{1.5cm}+L^{|\omega_0|}_{n_r}(\alpha_\rho \rho^2)
	\Bigl((2n_r'+|\omega_0'|-\alpha_\rho \rho^2)L^{|\omega_0'|}_{n_r'}(\alpha_\rho \rho^2)
	-2(n_r'+|\omega_0'|)L^{|\omega_0'|}_{n_r'-1}(\alpha_\rho \rho^2)\Bigr)
	\pm L^{|\omega_0|}_{n_r}(\alpha_\rho \rho^2) 
	L^{|\omega_0'|}_{n_r'}(\alpha_\rho \rho^2)\biggr]
	\nonumber \\
	&&\hspace{1cm}\times 
	H_{n_z}(\sqrt{\alpha_z}z) zH_{n_z'}(\sqrt{\alpha_z}z)\biggr\} \cos F(\rho,z)
	\delta_{K_3\mp\frac{1}{2} K_3'\pm\frac{1}{2}}\,.
\end{eqnarray}
\end{widetext}
\end{document}